\documentclass[reprint,superscriptaddress,amsmath,amssymb,aps,prc]{revtex4-1}

\usepackage{graphicx}
\usepackage{hyperref}
\usepackage{multirow,bigdelim}
\usepackage{siunitx}
\newcommand{\SIhyp}[2]{\SI[number-unit-product={\text{-}}]{#1}{#2}}
\DeclareSIUnit\clight{\text{\ensuremath{c}}}

\begin{document}

\title{In-beam $\gamma$-ray spectroscopy of $^{32}$Mg via direct reactions}

\author{N.~Kitamura}
\email{nkitamur@utk.edu}
\affiliation{Center for Nuclear Study, University of Tokyo, Wako, Saitama 351-0198, Japan}
\affiliation{Department of Physics and Astronomy, University of Tennessee, Knoxville, Tennessee 37996, USA}
\author{K.~Wimmer}
\affiliation{Instituto de Estructura de la Materia, CSIC, 28006 Madrid, Spain}
\affiliation{Department of Physics, University of Tokyo, Bunkyo, Tokyo 113-0033, Japan}
\affiliation{Department of Physics, Central Michigan University, Mt.\ Pleasant, Michigan 48859, USA}
\affiliation{National Superconducting Cyclotron Laboratory, Michigan State University, East Lansing, Michigan 48824, USA}
\author{T.~Miyagi}
\affiliation{TRIUMF, Vancouver, British Columbia V6T 2A3, Canada}
\author{A.~Poves}
\affiliation{Departamento de F\'isica Te\'orica and IFT UAM-CSIC, Universidad Aut\'onoma de Madrid, 28049 Madrid, Spain}
\author{N.~Shimizu}
\affiliation{Center for Nuclear Study, University of Tokyo, Wako, Saitama 351-0198, Japan}
\author{J.~A.~Tostevin}
\affiliation{Department of Physics, University of Surrey, Guildford, Surrey GU2 7XH, United Kingdom}
\author{V.~M.~Bader}
\affiliation{National Superconducting Cyclotron Laboratory, Michigan State University, East Lansing, Michigan 48824, USA}
\affiliation{Department of Physics and Astronomy, Michigan State University, East Lansing, Michigan 48824, USA}
\author{C.~Bancroft}
\affiliation{Department of Physics, Central Michigan University, Mt.\ Pleasant, Michigan 48859, USA}
\author{D.~Barofsky}
\affiliation{Department of Physics, Central Michigan University, Mt.\ Pleasant, Michigan 48859, USA}
\author{T.~Baugher}
\affiliation{National Superconducting Cyclotron Laboratory, Michigan State University, East Lansing, Michigan 48824, USA}
\affiliation{Department of Physics and Astronomy, Michigan State University, East Lansing, Michigan 48824, USA}
\author{D.~Bazin}
\affiliation{National Superconducting Cyclotron Laboratory, Michigan State University, East Lansing, Michigan 48824, USA}
\author{J.~S.~Berryman}
\affiliation{National Superconducting Cyclotron Laboratory, Michigan State University, East Lansing, Michigan 48824, USA}
\author{V.~Bildstein}
\affiliation{Department of Physics, University of Guelph, Guelph, Ontario N1G 2W1, Canada}
\author{A.~Gade}
\affiliation{National Superconducting Cyclotron Laboratory, Michigan State University, East Lansing, Michigan 48824, USA}
\affiliation{Department of Physics and Astronomy, Michigan State University, East Lansing, Michigan 48824, USA}
\author{N.~Imai}
\affiliation{Center for Nuclear Study, University of Tokyo, Wako, Saitama 351-0198, Japan}
\author{T.~Kr\"oll}
\affiliation{Institut f\"ur Kernphysik, Technische Universit\"at Darmstadt, 64289 Darmstadt, Germany}
\author{C.~Langer}
\affiliation{National Superconducting Cyclotron Laboratory, Michigan State University, East Lansing, Michigan 48824, USA}
\author{J.~Lloyd}
\affiliation{Department of Physics, Central Michigan University, Mt.\ Pleasant, Michigan 48859, USA}
\author{E.~Lunderberg}
\affiliation{National Superconducting Cyclotron Laboratory, Michigan State University, East Lansing, Michigan 48824, USA}
\affiliation{Department of Physics and Astronomy, Michigan State University, East Lansing, Michigan 48824, USA}
\author{F.~Nowacki}
\affiliation{Institut Pluridisciplinaire Hubert Curien, 67037 Strasbourg, France}
\author{G.~Perdikakis}
\affiliation{Department of Physics, Central Michigan University, Mt.\ Pleasant, Michigan 48859, USA}
\affiliation{National Superconducting Cyclotron Laboratory, Michigan State University, East Lansing, Michigan 48824, USA}
\author{F.~Recchia}
\affiliation{National Superconducting Cyclotron Laboratory, Michigan State University, East Lansing, Michigan 48824, USA}
\author{T.~Redpath}
\affiliation{Department of Physics, Central Michigan University, Mt.\ Pleasant, Michigan 48859, USA}
\author{S.~Saenz}
\affiliation{Department of Physics, Central Michigan University, Mt.\ Pleasant, Michigan 48859, USA}
\author{D.~Smalley}
\affiliation{National Superconducting Cyclotron Laboratory, Michigan State University, East Lansing, Michigan 48824, USA}
\author{S.~R.~Stroberg}
\affiliation{National Superconducting Cyclotron Laboratory, Michigan State University, East Lansing, Michigan 48824, USA}
\affiliation{Department of Physics and Astronomy, Michigan State University, East Lansing, Michigan 48824, USA}
\author{Y.~Utsuno}
\affiliation{Advanced Science Research Center, Japan Atomic Energy Agency, Tokai, Ibaraki 319-1195, Japan}
\affiliation{Center for Nuclear Study, University of Tokyo, Wako, Saitama 351-0198, Japan}
\author{D.~Weisshaar}
\affiliation{National Superconducting Cyclotron Laboratory, Michigan State University, East Lansing, Michigan 48824, USA}
\author{A.~Westerberg}
\affiliation{Department of Physics, Central Michigan University, Mt.\ Pleasant, Michigan 48859, USA}

\date{\today}

\begin{abstract}
\begin{description}
\item[Background] The nucleus $^{32}$Mg ($N=20$ and $Z=12$) plays a central role in the so-called ``island of inversion'' where in the ground states $sd$-shell neutrons are promoted to the $fp$-shell orbitals across the shell gap, resulting in the disappearance of the canonical neutron magic number $N=20$.
\item[Purpose] The primary goals of this work are to extend the level scheme of $^{32}$Mg, provide spin-parity assignments to excited states, and discuss the microscopic structure of each state through comparisons with theoretical calculations.
\item[Method] In-beam $\gamma$-ray spectroscopy of $^{32}$Mg was performed using two direct-reaction probes, one-neutron (two-proton) knockout reactions on $^{33}$Mg ($^{34}$Si). Final-state exclusive cross sections and parallel momentum distributions were extracted from the experimental data and compared with eikonal-based reaction model calculations combined with shell-model overlap functions.
\item[Results] Owing to the remarkable selectivity of the one-neutron and two-proton knockout reactions, a significantly updated level scheme for $^{32}$Mg, which exhibits negative-parity intruder and positive-parity normal states, was constructed. The experimental results were confronted with four different nuclear structure models.
\item[Conclusions] In some of these models, different aspects of $^{32}$Mg and the transition into the island of inversion are well described. However, unexplained discrepancies remain, and even with the help of these state-of-the-art theoretical approaches, the structure of this key nucleus is not yet fully captured.
\end{description}
\end{abstract}

\pacs{}
\keywords{}

\maketitle

\section{Introduction}

The breakdown of the canonical magic numbers in unstable nuclei has attracted much attention over the years~\cite{SOR08,NOW21}. To date, the disappearance of magicity has been reported in several regions of the nuclear chart away from the stability line. Such regions are sometimes associated with ``islands of inversion,'' in which deformed ground states are dominated by particle-hole ($n$p$n$h) excitations, resulting from diminished shell gaps and strong quadrupole correlations. The concept of an island of inversion was first applied for the neutron rich $sd$-shell nuclei around $N=20$~\cite{WAR90}. Later, other islands of inversion in the neutron-rich regions around $N=8$~\cite{SOR08}, $N=28$~\cite{CAU14}, $N=40$~\cite{LEN10}, and $N=50$~\cite{NOW16} have been proposed, thus revealing an archipelago in the ``sea'' of unstable nuclei. From a shell-model point of view, the primary driving force behind the islands of inversion is thought to be spin-isospin interactions between nucleons that dramatically alter the shell structure as a function of proton and neutron number~\cite{OTS01,OTS05,OTS10,OTS20}. These nucleon-nucleon correlations drive shell evolution and enable particle-hole excitations and promote deformation. However, the complete, universal understanding of the island-of-inversion physics is not yet obtained.

The nucleus $^{32}$Mg ($N=20$ and $Z=12$), regarded as the heart of the original island of inversion, has played the central role in the studies of evolving shell structure around the magic number $N=20$. The structure of this nucleus has been studied over the past several decades, and its low $2^+_1$~\cite{DET79} energy and high $B(E2;0^+_1\to2^+_1)$~\cite{MOT95,ELD21} value, which signify the breakdown of magicity, have been well established. To date, much effort has been devoted to a proper and unified description of the island of inversion~\cite{HIN11,CAU14,MAC16,MIY20}, and it continues to be an interesting subject to study experimentally~\cite{TRI08a,KAN10,WIM10,CRA16,ELD19}. To differentiate between various structural models, confrontation with additional, high-quality experimental data is required.

A variety of approaches have been employed for the spectroscopy of $^{32}$Mg. Since the late 1970s, $\beta$-$\gamma$ spectroscopy of $^{32}$Na has been performed~\cite{DET79,GUI84,KLO93}. In 2007, the $^{32}$Mg level scheme was extended by more detailed $\beta$-$\gamma$ measurements~\cite{MAT07,TRI08a}. Additionally, results from $\beta$-delayed neutron emission from $^{33}$Na have been reported~\cite{GUI84,NUM01}. Levels in $^{32}$Mg were also studied via nuclear reactions. These include proton inelastic scattering~\cite{TAK09}, secondary fragmentation of $^{46}$Ar~\cite{CRA16}, and Coulomb breakup of $^{33}$Mg~\cite{DAT16}. An inclusive measurement of the one-neutron knockout reaction from $^{33}$Mg was reported as well~\cite{KAN10}. A body of experimental information concerning $B(E2)$ values exists~\cite{MOT95,PRI99,IWA01,CHI01,CHU05,ELD21}. These measurements established the large collectivity of this nucleus with a strong deformation of the ground state. In 2010, the $0^+_2$ state was experimentally observed in the $t$($^{30}$Mg,$p$) reaction~\cite{WIM10}, and later the existence of this state was confirmed by an in-beam $\gamma$-ray spectroscopy experiment~\cite{ELD19}. From a simplified point of view, the $0^+_2$ state in $^{32}$Mg is interpreted as a counterpart of the near-spherical ground state of $^{30}$Mg~\cite{SCH09}. This $0_2^+$ can be understood in terms of a manifestation of shape coexistence, which is proposed to be a universal phenomenon in the islands of inversion~\cite{HEY11}. To first order, the competition of the two different structures and the emergence of the island of inversion can be explained by the subtle balance between the 2p2h intruder and 0p0h normal configurations. However, recent theoretical calculations have pointed out that the 4p4h configurations may also play an important role~\cite{CAU14,MAC16}.

In this paper, we report on a detailed spectroscopic study on $^{32}$Mg, using reaction probes with different sensitivities to the underlying nuclear structure. The one-neutron (two-proton) knockout reactions on $^{33}$Mg ($^{34}$Si), respectively sensitive to intruder and normal configurations, were exploited simultaneously. Our primary emphasis will be on negative-parity states, which provide us with valuable information on the neutron shell structure, and positive-parity states exclusively populated in two-proton knockout. The main results were presented in the Letter communication of Ref.~\cite{KIT21}. Here, we present additional information and details of the experimental results, and full comparisons with available theoretical models.

\section{Methods}

\subsection{Experimental details}

The experiment was performed at the National Superconducting Cyclotron Laboratory at Michigan State University. The experimental setup is identical to our earlier work on $\gamma$-ray spectroscopy of $^{30}$Mg~\cite{KIT20}, but with different magnetic-rigidity settings of the beamline and spectrograph. To produce secondary beams of $^{33}$Mg and $^{34}$Si via fragmentation reactions, a $^{48}$Ca beam at \SI{140}{MeV/nucleon} delivered from the Coupled Cyclotron Facility impinged on an \SIhyp{846}{mg/cm^2} thick $^9$Be production target. Reaction products were separated with the A1900 fragment separator~\cite{MOR03} employing a \SIhyp{300}{mg/cm^2} thick Al wedge degrader at the image plane. Data were collected in two separate settings optimized for the $^{33}$Mg and $^{34}$Si beams, respectively. The beams were directed onto a \SI{375}{mg/cm^2} thick $^{9}$Be secondary target at incident energies of \num{99.6} and \SI{94.8}{MeV/nucleon} for $^{33}$Mg and $^{34}$Si, respectively. The beams of interest had average on-target intensities of \num{620} and \SI{5.2e5}{s^{-1}}, and purities of \num{15} and \SI{66}{\percent}, respectively. Event-by-event identification of the incoming particles was performed using time-of-flight plastic scintillators installed in the beamline upstream of the secondary target.

The outgoing particles arising from the one-neutron (two-proton) knockout reactions on $^{33}$Mg ($^{34}$Si), induced by the secondary target, were momentum-analyzed by the S800 spectrograph~\cite{BAZ03a}. The standard S800 focal-plane detectors~\cite{YUR99}, i.e., a set of two cathode-readout drift chambers, an ionization chamber with segmented anodes, and a plastic scintillator, installed in the focal-plane box of the S800 spectrograph, provided unambiguous identification of the outgoing particles by energy loss and time-of-flight measurements as well as reconstruction of their momenta. A set of parallel-plate avalanche counters, installed at the intermediate plane upstream of the secondary target, was used to track the incoming ions. This enabled us to correct for the momentum of the incoming particle and improve the resolution for the momentum distributions of the outgoing ions. The overall resolutions were \SI{0.08}{GeV/\clight} (\SI{0.26}{GeV/\clight}) for the $^{33}$Mg ($^{34}$Si) settings, dominated by the element-specific energy loss of the beam and reaction products in the target.

The secondary target was surrounded by the state-of-the-art array of $\gamma$-ray detectors, Gamma-Ray Energy Tracking In-beam Nuclear Array, GRETINA~\cite{PAS13,FAL16,WEI17}. At the time of the experiment, the array consisted of seven modules and it was set up with four modules at \SI{58}{\degree} polar angle and the remaining three at \SI{90}{\degree} with respect to the beam axis. Each GRETINA module houses four high-purity germanium crystals. Combining the 36-fold electrical segmentation of the crystal and dedicated online waveform decomposition algorithms, $\gamma$-ray hit-position reconstruction with a sub-segment resolution was accomplished. Together with the reconstructed velocity and angle of the outgoing particle at the target position, the $\gamma$-ray position information was used for the Doppler correction on an event-by-event basis. For the identification of $\gamma$-ray peaks, the add-back procedure based on the nearest-neighbor algorithm~\cite{WEI17} was employed for an improved peak-to-total ratio.

\subsection{Reaction model calculations}

One-neutron knockout reactions have been well established as a powerful spectroscopic tool for unstable nuclei~\cite{HAN03,GAD08}. Knockout reactions involving the removal of two protons from neutron-rich nuclei have also been shown to proceed as direct reactions, and theoretical prescriptions to model such reactions have already been documented in Refs.~\cite{TOS04,TOS06,SIM09a,SIM09b}. In the reactions of this kind, the parallel momentum distribution of the knockout residue is indicative of the total angular momentum of the removed nucleons~\cite{BAZ03b,YON06,SAN11,LON20}, and thus the experimental momentum distribution associated with each final state can be used to infer the spin and parity.

The experimental cross sections populating individual states can be compared with theoretical predictions. For one-neutron knockout reactions, by using the reaction theory based on the eikonal and sudden approximations, the theoretical cross section for a final state, $J_f^\pi$, is computed as the sum of contributions from each single-particle orbital with quantum numbers $nlj$
\begin{equation}
\label{eq:theoxsec}
\sigma_\mathrm{th} = \sum_{nlj} \left(\frac{A}{A-1}\right)^N C^2S(J_f^\pi,nlj) \sigma_\mathrm{sp} \left[S_n+E_\mathrm{x}(J_f^\pi),nlj\right].
\end{equation}
$C^2S$ is the shell-model spectroscopic factor, i.e., the norm of the one-neutron overlap function, which contains nuclear structure information. Details of the structural calculation are given in the next section. The first coefficient, $[A/(A-1)]^N$, with $N$ being the major oscillator quantum number, is the center-of-mass correction factor~\cite{DIE74} to shell-model spectroscopic factors. $\sigma_\mathrm{sp}$ is the single-particle cross section, evaluated at the effective neutron separation energy $S_n+E_\mathrm{x}(J_f^\pi)$, and is taken from the reaction model calculations. In this work, the approach described in Ref.~\cite{GAD08} was adopted. Theoretical momentum distributions can be calculated in the same framework as for the cross sections. Generally, the higher the angular momentum is transferred and the more bound the removed nucleon, the broader the momentum distribution becomes. Likewise, theoretical two-proton knockout cross sections and momentum distributions can be calculated using the eikonal-based reaction theory. In this case, two-nucleon amplitudes (TNAs) are required as input, not only for cross sections but also for momentum distributions.

\subsection{Structural calculations}
\label{sect:structuralcalc}

In this work, four different shell-model interactions are employed for the calculation of final-state excitation energies, spectroscopic factors, and TNAs. Details of each are given in the following.

The SDPF-M interaction~\cite{UTS99} was developed in 1999 and is now regarded as a traditional interaction for neutron-rich nuclei at and around the island of inversion. The model space includes the full $sd$ shell and the lower half of the $fp$ shell, i.e., the $1f_{7/2}$ and $2p_{3/2}$ orbitals, for both neutrons and protons. In the calculations using this interaction, all states with $J\leq 5$ below the neutron threshold were calculated without any truncation. This allowed for the calculation of the theoretical inclusive cross section.

SDPF-U-MIX is one of the state-of-the-art shell-model interactions. It has been shown to provide excellent reproductions of level energies of $^{30}$Mg, $^{32}$Mg, and $^{34}$Si~\cite{CAU14}. Compared to SDPF-M, the model space is extended to include the full neutron $fp$ shell, while the protons are confined in the $sd$ shell. Because of computational limitations, calculations were performed with a 6p6h (5p5h) truncation for positive-parity (negative-parity) states. The number of calculated levels was limited such that the inclusive cross section to all bound shell-model final states cannot be computed. Therefore, the summed cross section presented in Fig.~\ref{fig:xsec1n} should be considered as a lower limit for the theoretical inclusive cross section.

Another state-of-the-art interaction, EEdf1~\cite{TSU17,TSU20}, uses the extended Kuo-Krenciglowa method~\cite{TSU14} and the Entem-Machleidt QCD-based nucleon-nucleon interaction~\cite{ENT03} to microscopically derive the shell-model two-body matrix elements. The single-particle energies are empirically determined through a fit to selected experimental observables. The model space of this interaction comprises of the full $sdpf$ shells for both neutrons and protons, which is computationally demanding. Practical shell-model calculations were performed utilizing the KSHELL code~\cite{SHI19}. A truncation at 6p6h (7p7h) was imposed for positive-parity (negative-parity) states, and only the lowest six (three) states for $0^+$, $2^+$, and $4^+$ ($1^-$, $2^-$, and $3^-$) have been calculated in this work.

A shell-model interaction generated using the valence-space in-medium similarity renormalization group~\cite{MIY20}, dubbed IMSRG in the present work, is used for comparisons as well. For the derivation of the shell-model Hamiltonian, the 1.8/2.0 (EM) nucleon-nucleon plus three-nucleon interaction~\cite{HEB11}, derived based on the chiral effective-field theory, was used as input. The IMSRG evolution was performed in the 13 major-shell space with the spherical harmonic oscillator (HO) basis at $\hbar\omega=16$ MeV. An additional truncation $E_\mathrm{3max}$, defined as the sum of the three-nucleon HO principal quantum numbers, was introduced for the three-nucleon interaction. We used $E_\mathrm{3max}=16$ in this study. A center-of-mass parameter of $\beta=3$, determined from the convergence of calculated level energies, was adopted for the valence-space diagonalization (see Ref.~\cite{MIY20} for details). The model space was taken to be the $sd$ shell for protons and the $sd$ plus the lower half of the $fp$ shell for neutrons. This small model space allowed us to calculate all states with $J \leq 5$ below the neutron threshold. For the calculation of spectroscopic factors and TNAs, presently bare annihilation operators are used. The operators should ideally be evolved consistently. We note that the consistent evolution changes the spectroscopic factors by roughly \SI{10}{\percent}~\cite{STO21}. Since the TNA operator involves the two annihilation (or creation) operators similarly to the spectroscopic factor, a na\"ive estimation provides the same size of renormalization effect for the TNAs.

\section{Analysis and experimental results}

\subsection{Level scheme}

\begin{figure*}[tb]
\includegraphics[scale=0.5]{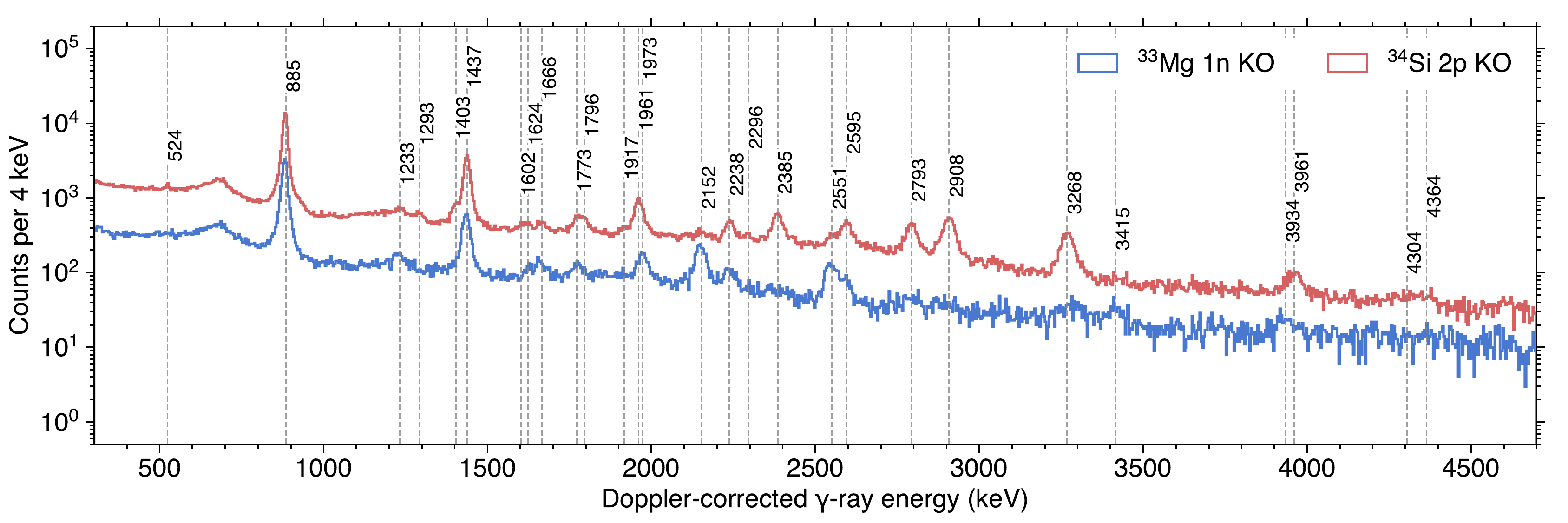}
\caption{Doppler-corrected add-back $\gamma$-ray spectra from $^{32}$Mg recorded in coincidence with the incoming projectiles of $^{33}$Mg (blue) and $^{34}$Si (red). Identified peaks are indicated by vertical dashed lines together with their transition energies, regardless of whether the transition was placed in the level scheme or not.}
\label{fig:gamma}
\end{figure*}

The level scheme of $^{32}$Mg was constructed based on $\gamma$ rays emitted from $^{32}$Mg and their $\gamma$-$\gamma$ coincidence relations. Doppler-corrected $\gamma$-ray spectra from the two reactions are shown in Fig.~\ref{fig:gamma}. The most prominent peak seen in both histograms is the \SIhyp{885}{keV} $2^+_1\to0^+_1$ transition. Because of the lifetime of the $2^+_1$ state, the $\gamma$-ray emission takes place at a slower velocity than that at the reaction point, resulting in a distortion of the Doppler-corrected $\gamma$-ray lineshape. The known, experimental $B(E2)$ value of this transition translates to a half-life of \SI{11.4(20)}{ps}~\cite{OUE11}. The observed peak centroid is shifted to a lower energy of \SI{882}{keV}, which agrees with a simulated lineshape for the delayed emission based on this lifetime. Hereafter, a value of \SI{885}{keV}~\cite{OUE11} is used as the $2^+_1\to0^+_1$ transition energy.

The second highest peak in the spectra at \SI{1437}{keV} corresponds to the $4^+_1\to2^+_1$ transition~\cite{TAK09}. The $4^+_1$ state is expected to have a short lifetime of around \SI{1}{ps}, because of its high collectivity, as inferred from theoretical calculations. This was recently verified experimentally~\cite{ELD21}. We note that, however, such a short lifetime is beyond the sensitivity of the present setup. In the present analysis, all peaks, except the \SI{885}{keV} transition, are assumed to be prompt. Since the present experiment was not designed to efficiently observe $\gamma$ rays from long-lived states (typically more than \SI{1}{ns}), the $\gamma$ ray from the $0_2^+$ state decaying to $2^+_1$, with a transition energy of \SI{172}{keV}, was not observed, unlike the measurement in Ref.~\cite{ELD19}, where the experiment was optimized to observe such transitions.

\begin{figure}[tb]
\includegraphics[scale=0.5]{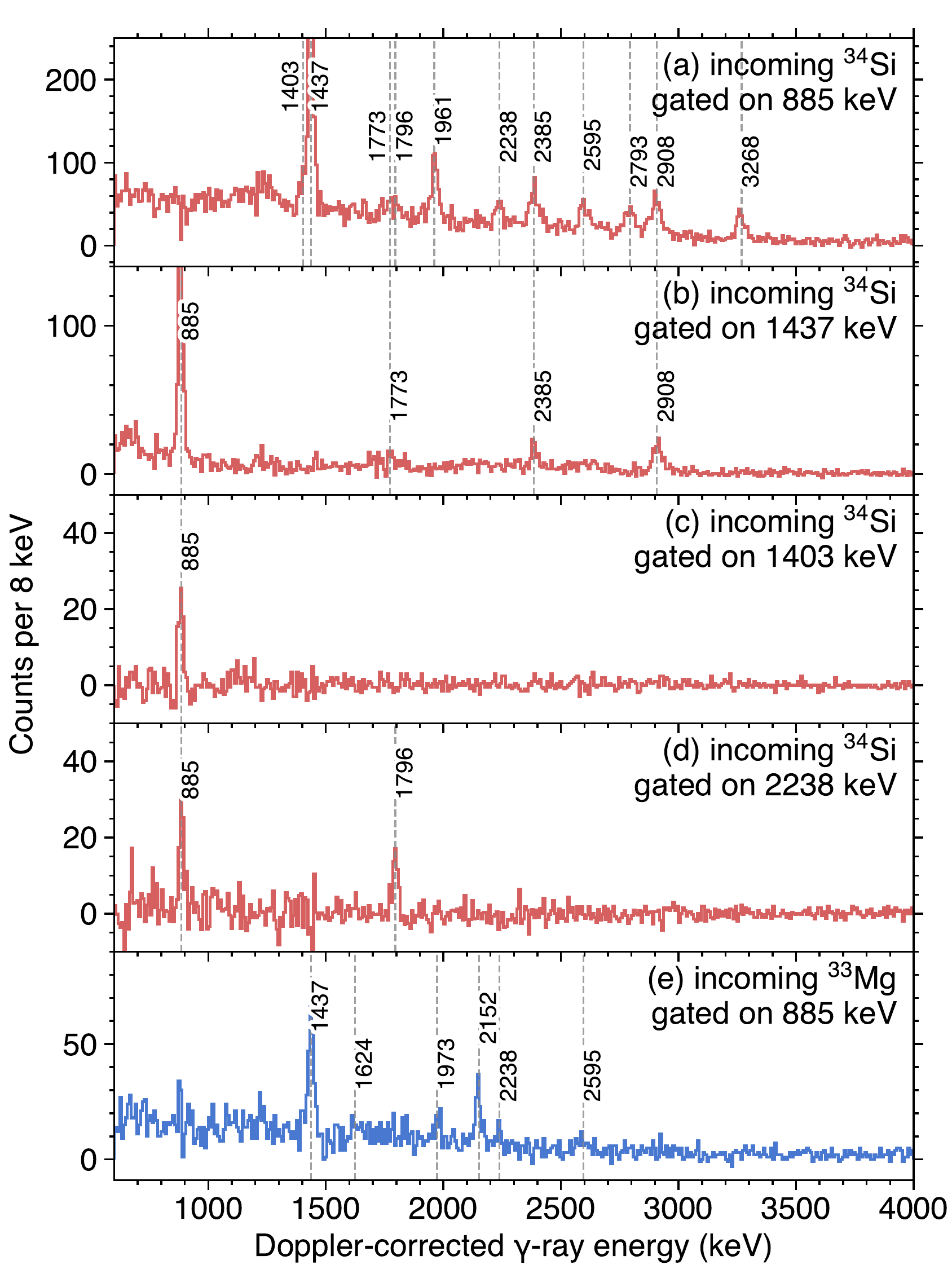}
\caption{Doppler-corrected add-back $\gamma$-$\gamma$ coincidence spectra from (a--d) the two-proton knockout and (e) one-neutron knockout reactions. Background contributions have been subtracted. Identified coincidence lines are indicated by vertical dashed lines with their transition energies.}
\label{fig:gamgam}
\end{figure}

\begin{figure*}[tb]
\includegraphics[scale=0.5]{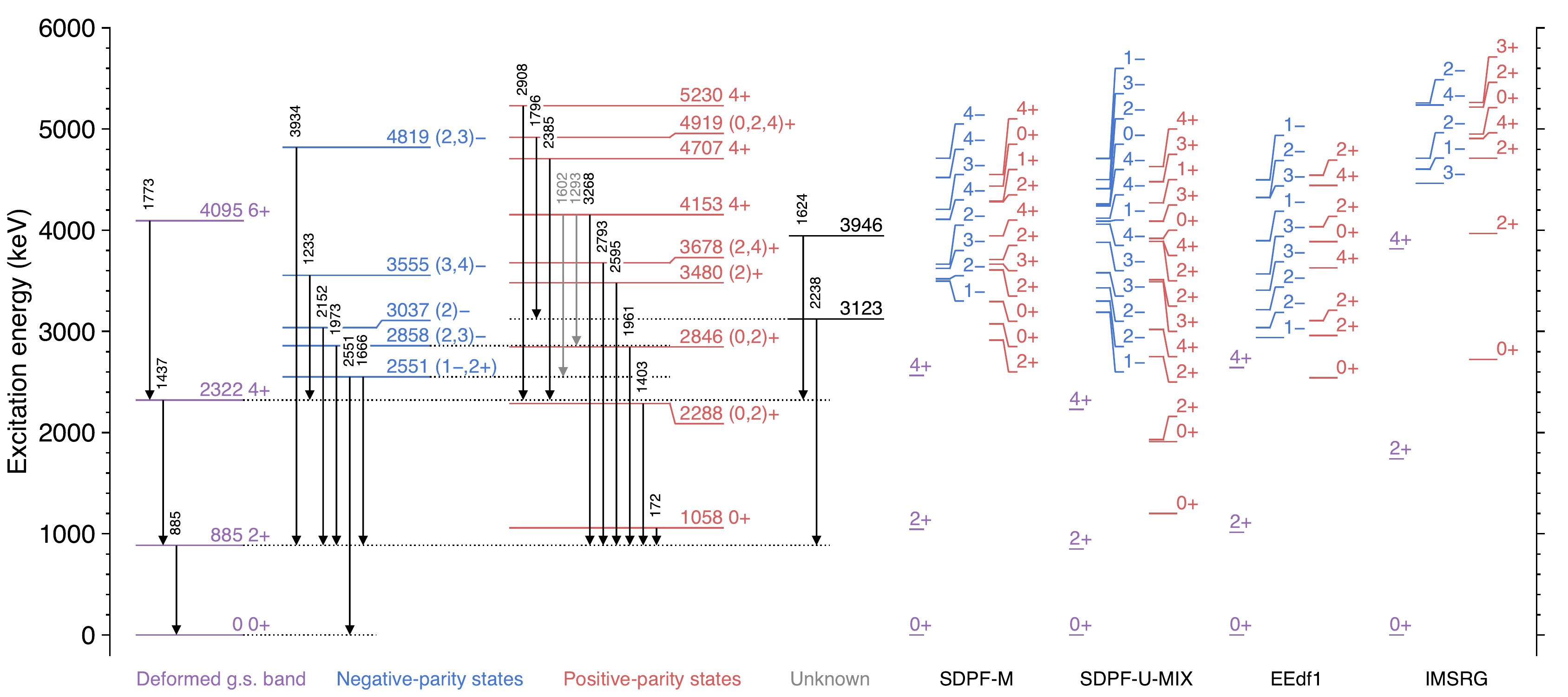}
\caption{Level scheme of $^{32}$Mg constructed in the present work. Placed transitions are indicated by the arrows. Tentative placements are shown in gray. States are sorted into four categories: deformed ground-state band, negative- and positive-parity states (including candidates), and states with unknown properties. The experimental level scheme is compared with theoretical calculations using the SDPF-M, SDPF-U-MIX, and EEdf1 shell-model interactions. Additionally, shell-model results using an interaction derived from the IMSRG approach are shown.}
\label{fig:levelscheme}
\end{figure*}

The high statistics collected in the present measurement allowed us to investigate $\gamma$-$\gamma$ coincidence relationships. The $\gamma$-ray spectra from the $\gamma$-$\gamma$ analysis are shown in Fig.~\ref{fig:gamgam}. Background-subtracted spectra were generated in the standard way by subtracting a background cut taken at slightly higher energy with respect to the energy of interest, normalized to the cut window width. An updated level scheme for $^{32}$Mg, constructed in the present analysis, is presented in Fig.~\ref{fig:levelscheme}.

The \num{1403}-, \num{1961}-, \num{2793}-, \num{3268}-, \num{2385}-, \num{1796}-, and \SIhyp{2908}{keV} transitions were newly placed in the level scheme, based on Figs.~\ref{fig:gamgam}(a--d). One of the notable results from the coincidence analysis is that the \SIhyp{1403}{keV} transition was found to be in coincidence with the \SIhyp{885}{keV} transition. As evidenced by the $\gamma$-ray spectrum of Fig.~\ref{fig:gamgam}(c), no other transitions are in coincidence with the \SIhyp{1403}{keV} $\gamma$ ray. Following this observation, a new state at \SI{2288}{keV}, built on top of the $2_1^+$ state, was established. This level is located slightly lower than the $4_1^+$ state at \SI{2322}{keV}, which seems to be irregular. A simulation was carried out to explore the possibility of having a \num{1403}-\num{1437}-\num{885} $\gamma$-ray cascade. The result showed that a clear peak at \SI{1437}{keV} would be observed in the spectrum of Fig.~\ref{fig:gamgam}(c) if the \SI{1403}{keV} $\gamma$ ray fed the $4_1^+$ state, thus excluding this possibility. It was also found that the peak observed at \SI{1961}{keV} in the two-proton knockout spectrum is different from the \SIhyp{1973}{keV} transition seen in one-neutron knockout, which corresponds to the \SIhyp{1972.9(5)}{keV} line established in $\beta$-decay studies~\cite{MAT07,TRI08a,OUE11}. The \SIhyp{1961}{keV} transition was found to feed the $2^+_1$ \SIhyp{885}{keV} state, thus a new state at \SI{2846}{keV} was established. We emphasize that the good energy resolution enabled by GRETINA was particularly beneficial for the construction of the updated level scheme, otherwise the closely-spaced $\gamma$ lines, i.e., \num{1403}-\SI{1437}{keV}, and \num{1961}-\SI{1973}{keV}, would not have been disentangled.

We note that, for some of the transitions with limited statistics, the placements were guided by those made in previous studies. The \num{1666}-\num{885}, \num{3934}-\num{885}, and \num{1232}-\num{1437} $\gamma$-ray cascades were previously observed in $\beta$-decay experiments~\cite{MAT07,TRI08a} and therefore the placement of these transitions was adopted. A $\gamma$-ray energy doublet at \SI{1232}{keV} was proposed in Ref.~\cite{MAT07}, in which the state at \SI{4785}{keV} decays to the \SIhyp{3553}{keV} state (\SI{3555}{keV} in the present level scheme), emitting a \SIhyp{1232}{keV} $\gamma$ ray, and this state further decays to the $4^+_1$ state again emitting a \SIhyp{1232}{keV} $\gamma$ ray. However, in the present level scheme, the existence of the state at \SI{4785}{keV} is not assumed.

The \SIhyp{2238}{keV} $\gamma$-ray is in coincidence with the \num{1796}- and \SIhyp{885}{keV} transitions, as can be seen in Fig.~\ref{fig:gamgam}(d), and this transition was placed on top of the \SIhyp{885}{keV} state. The ordering of these transitions was inferred from their intensities, i.e., the \SIhyp{2238}{keV} transition is more intense than \SI{1796}{keV}. This ordering is supported also by the previous measurement of proton inelastic scattering~\cite{TAK09}, where only the \SIhyp{2238}{keV} transition is observed and placed on top of the \SIhyp{885}{keV} state.

The \SIhyp{1773}{keV} transition is assigned as the $6^+\to4^+$ transition in Ref.~\cite{CRA16}. A hint of a peak, which may correspond to this transition, was observed in the singles and coincidence spectra.

In the present analysis, the existence of three closely-lying $\gamma$-ray peaks is suggested around \SI{1600}{keV} in the singles spectra. In $\beta$-decay measurements~\cite{MAT07,TRI08a}, the \SI{1666}{keV} transition was reported to compete with the \SI{2551}{keV} ground-state transition. It is worthwhile to see if the branching ratio of the two transitions from the \SIhyp{2551}{keV} state is consistent with those measured previously. In the two-proton knockout spectrum, the intensity for \SI{1666}{keV} is higher than \SI{2551}{keV} (see Table~\ref{tab:gamint}), conflicting with the $\beta$-decay results, where the relative intensity of the \SI{1666}{keV} transition with respect to \SI{2551}{keV} was reported to be around \SI{30}{\percent}~\cite{MAT07,TRI08a}. It is therefore likely that the \SIhyp{1666}{keV} transition in Fig.~\ref{fig:gamma} is contaminated by yet another transition. Nevertheless, in the one-neutron knockout spectrum, the intensity for \SI{1666}{keV} is \SI{38(4)}{\percent} relative to \SI{2551}{keV} and is consistent with the $\beta$-decay result.

The \SIhyp{1602}{keV} transition is compatible with the spacing between the \num{4153}- and \SIhyp{2551}{keV} levels (shown in gray in Fig.~\ref{fig:levelscheme}). Likewise, the \SIhyp{1293}{keV} transition agrees with the spacing between \num{4153}- and \SIhyp{2858}{keV} states. These transitions are only tentatively placed in the level scheme and are not used for constraining spin-parity assignments.

\begin{table}[tb]
\centering
\caption{Observed $\gamma$-ray energies and their placements. If observed, relative intensities, with respect to the \SIhyp{885}{keV} transition, are indicated. The energies are given in units of \si{keV}. The quoted uncertainties for the relative intensities include the statistical contribution only. Tentative placements are indicated by parentheses in the second column.}
\label{tab:gamint}
\begin{tabular}{S[table-align-text-post=false]cSS}
\toprule
{Transition energy} & {Placement} & {$2p$ KO} & {$1n$ KO} \\
\colrule
524(1)              &                 &   0.9(1)  &          \\
885\footnotemark[1] & 885$\to$0       & 100.0(4)  & 100.0(8) \\
1233(2)             & 3555$\to$2322   &   0.8(2)  &   2.4(4) \\
1293(3)             & (4153$\to$2858) &   1.1(2)  &          \\
1403(2)             & 2288$\to$885    &   3.5(2)  &          \\
1437(2)             & 2322$\to$885    &  38.4(4)  &  25.0(6) \\
1602(4)             & (4153$\to$2551) &   0.8(2)  &          \\
1624(3)             & 3946$\to$2322   &   1.3(2)  &   1.3(4) \\
1666(3)             & 2551$\to$885    &   1.9(2)  &   3.8(4) \\
1773(3)             & 4095$\to$2322   &   2.7(2)  &   2.9(5) \\
1796(3)             & 4919$\to$3123   &   3.4(2)  &   0.8(5) \\
1917(4)             &                 &   1.8(2)  &          \\
1961(3)             & 2846$\to$885    &  11.5(5)  &          \\
1973(3)             & 2858$\to$885    &   1.8(4)  &   7.7(5) \\
2152(3)             & 3037$\to$885    &   1.4(2)  &  16.5(6) \\
2238(3)             & 3123$\to$885    &   5.3(2)  &   5.0(5) \\
2296(6)             &                 &   1.5(2)  &   0.7(4) \\
2385(4)             & 4707$\to$2322   &   9.4(3)  &   0.9(4) \\
2551(4)             & 2551$\to$0      &   1.8(2)  &  10.0(6) \\
2595(4)             & 3480$\to$885    &   6.6(3)  &   3.3(5) \\
2793(4)             & 3678$\to$885    &   8.1(3)  &   1.9(4) \\
2908(4)             & 5230$\to$2322   &  13.7(3)  &   1.9(4) \\
3268(5)             & 4153$\to$885    &  10.8(2)  &   2.2(4) \\
3415(7)             &                 &   0.7(2)  &   1.9(4) \\
3934(8)             & 4819$\to$885    &           &   1.1(4) \\
3961(7)             &                 &   1.9(2)  &          \\
4304(20)            &                 &   0.5(2)  &          \\
4364(13)            &                 &   1.0(2)  &          \\
\botrule
\end{tabular}
\footnotetext[1]{\SI{885.3(1)}{keV} according to Ref.~\cite{OUE11}.}
\end{table}

\subsection{Cross sections}

For the one-neutron knockout reaction, the inclusive cross section, i.e., the sum of cross sections populating all the bound states of $^{32}$Mg, was measured to be $\sigma_{1n}^\mathrm{incl} = \SI{104(5)}{mb}$, with the fluctuation in the secondary beam composition being the leading source of the systematic uncertainty. The present value shows good agreement with those measured in this mass region~\cite{TER08,KIT20}. However, it is larger than that given in the previous inclusive one-neutron knockout measurement of Ref.~\cite{KAN10}, in which the cross section was reported to be \SI{74(4)}{mb}. The reaction was performed at a high incident energy, \SI{898}{MeV/nucleon}, and with a carbon reaction target, and thus these two cross sections cannot be compared directly. The reaction model calculation coupled with shell-model spectroscopic factors can be used to predict the inclusive cross section by summing the exclusive cross sections given by Eq.~(\ref{eq:theoxsec}). With the SDPF-M interaction, the theoretical inclusive cross section was calculated to be \SI{107(10)}{mb}. As discussed in Sec.~\ref{sect:structuralcalc}, with the other effective interactions not all bound states could be calculated, and the theoretical cross sections are thus just lower limits. From the systematics, a quenching of the experimental cross section with respect to the theory prediction of \num{0.91(10)}~\cite{TOS14,TOS21} is expected. The measured cross section agrees with the prediction within the stated uncertainty. We note that, as was done for the theoretical inclusive cross section for the one-neutron knockout reaction from $^{31}$Mg~\cite{KIT20}, the theoretical uncertainty was estimated by artificially varying the neutron threshold ($S_n = \SI{5778}{keV}$~\cite{OUE11}) by \SI{\pm 500}{keV} when taking the sum. This large uncertainty implies that sizable strengths are distributed around the threshold.

The measured inclusive cross section for the two-proton knockout reaction is $\sigma_{2p}^\mathrm{incl} = \SI{0.96(8)}{mb}$. The present value compares well with those measured previously, even though it tends to be slightly higher than \num{0.76(10)}~\cite{BAZ03b} and \SI{0.86(8)}{mb}~\cite{ZWA05}. Comparison with theoretical calculations will be detailed later in Sec.~\ref{sect:comparisontosm}.

\begin{figure*}[tb]
\includegraphics[scale=0.5]{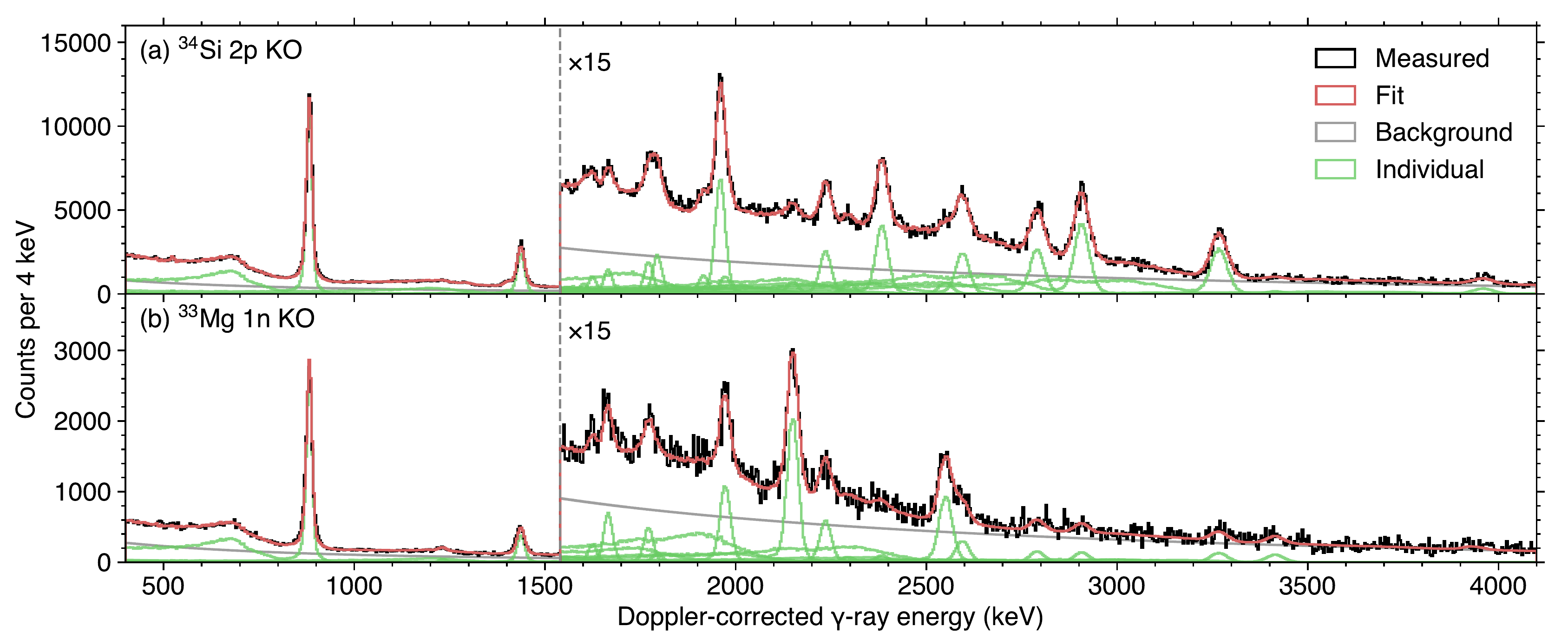}
\caption{Fitting results for the $\gamma$-ray spectra obtained from the (a) two-proton and (b) one-neutron knockout reactions. The black histogram shows the experimental Doppler-corrected spectrum without add-back. The fit function is shown in the red histogram. The green and gray histograms respectively represent the individual response functions and the background contribution. For the response function of the \SIhyp{885}{keV} $\gamma$ ray, the known lifetime~\cite{OUE11} was taken into account.}
\label{fig:fit}
\end{figure*}

In order to deduce the cross sections leading to each individual state, $\gamma$-ray intensities need to be extracted from the experimental $\gamma$-ray spectra. GRETINA response functions, generated by a dedicated Monte Carlo simulation package~\cite{RIL21}, were used to fit the observed spectra. In the fitting process, a smooth, double-exponential curve was added to the fit function to account for the continuous prompt $\gamma$-ray background, which is often observed in in-beam measurements (see, for example, Ref.~\cite{STO14}). The fit results to the $\gamma$-ray spectra from the two-proton and one-neutron knockout reactions are shown in Figs.~\ref{fig:fit}(a) and (b), respectively. The relative $\gamma$-ray intensities extracted from the fits are summarized in Table~\ref{tab:gamint}. Exclusive cross sections, obtained by balancing the $\gamma$-ray intensities, are tabulated in Table~\ref{tab:values}. It should be noted that the \SIhyp{172}{keV} transition~\cite{WIM10,ELD19} is delayed, likely having a half-life of more than \SI{1}{ns}, and in the present measurement this transition did not present any experimentally-observable structure in the $\gamma$-ray spectra. Therefore, the cross section populating the $0^+_2$ state is always included in the ground-state cross section.

\begin{table*}[tb]
\caption{Experimental cross sections for the two-proton ($\sigma_{2p}$) and one-neutron ($\sigma_{1n}$) knockout reactions. The uncertainties include systematic contributions propagating from the inclusive cross section and the $\gamma$-ray detection efficiency, which is estimated to be \SI{3}{\percent}. Cross-section ratios ($r$), defined as Eq.~(\ref{eq:ratio}), are indicated as well. Single-particle orbitals ($nlj$), their theoretical single-particle cross sections ($\sigma_\mathrm{sp}$), and the resulting spectroscopic factors ($C^2S_\mathrm{exp}$) are presented together. $\log ft$ values in the $\beta$ decay of $^{33}$Na, taken from Ref.~\cite{TRI08a}, are also shown.}
\label{tab:values}
\begin{tabular}{S[table-format=4.0,table-align-text-post=false]lSSS[retain-explicit-plus]llSSS}
\toprule
{$E_\mathrm{x}$ (keV)} & {$J^\pi$} & {$\sigma_{2p}$ (mb)} & {$\sigma_{1n}$ (mb)} & {$r$} & {} & {$nlj$} & {$\sigma_\text{sp}$ (mb)} & {$C^2S_\mathrm{exp}$} & {$\log ft$} \\
\colrule
   0        & $0^+$       & 0.275(30) & 13.0(29) & +0.40 &                        & $2p_{3/2}$                 & 39.3  & 0.33(7) &        \\
 885        & $2^+$       & 0.072( 9) & 27.7(21) & -0.55 &                        & $2p_{3/2}$ \SI{69}{\percent} (\SI{19.1(22)}{mb}) & 32.3  & 0.59(7) &        \\
            &             &           &          &       &                        & $1f_{7/2}$ \SI{31}{\percent} (\SI{8.6(20)}{mb})  & 19.4  & 0.44(5) &        \\
2288        & $(0,2)^+$   & 0.023( 2) &          & +1    &                        &                            &       &         &        \\
2322        & $4^+$       & 0.071( 7) & 12.9(12) & -0.25 &                        & $1f_{7/2}$                 & 17.2  & 0.75(6) &        \\
2551        & $(1^-,2^+)$ & 0.020( 3) & 11.4( 9) & -0.68 & \ldelim\{{2}{*}[$1^-$] & $2s_{1/2}$ \SI{45}{\percent} (\SI{5.1(12)}{mb})  & 29.3  & 0.18(4) &        \\
            &             &           &          &       &                        & $1d_{3/2}$ \SI{55}{\percent} (\SI{6.3(15)}{mb})  & 18.0  & 0.35(8) &        \\
            &             &           &          &       & \ldelim\{{2}{*}[$2^+$] & $2p_{3/2}$ \SI{92}{\percent} (\SI{10.5(12)}{mb}) & 24.2  & 0.44(5) &        \\
            &             &           &          &       &                        & $1f_{7/2}$ \SI{8}{\percent}  (\SI{0.9(9)}{mb})   & 16.9  & 0.05(5) &        \\
2846        & $(0,2)^+$   & 0.077( 7) &          & +1    &                        &                            &       &         &        \\
2858        & $(2,3)^-$   & 0.005( 3) &  6.4( 6) & -0.85 &                        & $2s_{1/2}$\footnotemark[1] & 28.0  & 0.23(2) & 5.8(2) \\
            &             &           &          &       &                        & $1d_{3/2}$\footnotemark[1] & 17.5  & 0.37(3) &        \\
3037        & $(2)^-$     & 0.009( 2) & 13.7( 9) & -0.86 &                        & $2s_{1/2}$ \SI{59}{\percent} (\SI{8.1(11)}{mb})  & 27.3  & 0.30(4) & 5.0(1) \\
            &             &           &          &       &                        & $1d_{3/2}$ \SI{41}{\percent} (\SI{5.6(10)}{mb})  & 17.2  & 0.33(6) &        \\
3123        &             & 0.013( 3) &  3.5( 6) & -0.43 &                        &                            &       &         &        \\
3480        & $(2)^+$     & 0.044( 4) &  2.7( 4) & +0.27 &                        &                            &       &         &        \\
3555        & $(3,4)^-$   & 0.006( 1) &  2.0( 3) & -0.53 &                        & $1d_{3/2}$                 & 16.4  & 0.12(2) & 5.4(1) \\
3678        & $(2,4)^+$   & 0.054( 5) &  1.6( 4) & +0.57 &                        &                            &       &         &        \\
3946        &             & 0.009( 2) &  1.1( 3) & -0.08 &                        &                            &       &         &        \\
4095        & $6^+$       & 0.018( 2) &  2.4( 4) & -0.09 &                        &                            &       &         &        \\
4153        & $4^+$       & 0.084( 6) &  1.8( 4) & +0.67 &                        &                            &       &         &        \\
4707        & $4^+$       & 0.063( 6) &  0.8( 4) & +0.80 &                        &                            &       &         &        \\
4819        & $(2,3)^-$   &           &  0.9( 3) & -1    &                        & $2s_{1/2}$\footnotemark[1] & 21.9  & 0.04(1) & 4.9(2) \\
            &             &           &          &       &                        & $1d_{3/2}$\footnotemark[1] & 14.8  & 0.06(2) &        \\
4919        & $(0,2,4)^+$ & 0.023( 2) &  0.7( 4) & +0.58 &                        &                            &       &         &        \\
5230        & $4^+$       & 0.091( 8) &  1.6( 4) & +0.73 &                        &                            &       &         &        \\
{Inclusive} &             & 0.96(8)   &  104( 5) &       &                        &                            &       &         &        \\
{Unplaced}  &             & 0.056( 6) &  2.1( 5) &       &                        &                            &       &         &        \\
\botrule
\end{tabular}
\footnotetext[1]{Due to the lack in statistics, the $2s_{1/2}$ and $1d_{3/2}$ components could not be reliably determined from the fit to the momentum distribution. The spectroscopic factors are calculated assuming two extreme cases, pure $2s_{1/2}$ or $1d_{3/2}$.}
\end{table*}

In Fig.~\ref{fig:gamma}, it can be seen that the \num{1973}-, \num{2152}-, and \SIhyp{2551}{keV} transitions are enhanced in the one-neutron knockout spectrum, as compared to two-proton knockout. These transitions correspond to the de-excitation of the states at \num{2858}, \num{3037}, and \SI{2551}{keV}, respectively. These three states have large cross sections of more than \SI{5}{mb}, and from a simple perspective, these states can be interpreted as intruder-dominated ones, implying the large overlaps with the ground state of $^{33}$Mg. It is worthwhile to compare this observation with $\beta$-decay results. The states at \num{2858} and \SI{3037}{keV} are populated in $\beta$ decay of $^{32}$Na~\cite{TRI08a}, making them candidates for negative-parity states, assuming a spin-parity of $(3^-)$ for the $^{32}$Na ground state. Direct measurement of the $^{32}$Na ground-state $J^\pi$ has not been reported yet. The ground-state $J^\pi$ was assigned as $(3^-,4^-)$ in Ref.~\cite{KLO93}, but this was guided by shell-model calculations and should be seen as a tentative one. The $(3^-)$ assumption made in Ref.~\cite{TRI08a} was again guided by shell-model calculations and no $\beta$-decay branches to $0^+_1$, $2^+_1$, and $4^+_1$. The $\log ft$ values associated with the decay to the 2858- and \SIhyp{3037}{keV} states are low ($< 6.0$) and compatible with allowed transitions~\cite{TRI08a}. The state at \SI{2551}{keV} was also observed in $\beta$ decay, but populated only via the feeding from higher-lying states, and is a candidate for either $2^+$ or $1^-$. In Ref.~\cite{TRI08a}, the associated $\log ft$ of this state is large and a $2^+$ assignment was proposed. Nevertheless, a $1^-$ assignment is still possible, if one assumes a forbidden transition.

Conversely, in Fig.~\ref{fig:gamma}, it is seen that some of the transitions observed in the two-proton knockout spectrum are enhanced as compared to one-neutron knockout.  More specifically, $\gamma$-ray transitions of \num{1403}, \num{1796}, \num{1961}, \num{2385}, \num{2793}, \num{2908}, and \SI{3268}{keV} are enhanced or observed only in the two-proton knockout reaction. These $\gamma$ rays correspond to the de-excitation of the states at \num{2288}, \num{4919}, \num{2846}, \num{4707}, \num{3678}, \num{5230}, and \SI{4153}{keV}, respectively. Those states are strongly populated with cross sections of more than \SI{0.02}{mb}. From these observations, the seven states mentioned above are considered to originate from proton excitations and are therefore presumably of normal nature.  As will be discussed later, states strongly populated in the two-proton knockout reaction predominantly have spin-parities of either $0^+$, $2^+$, or $4^+$ (see Fig.~\ref{fig:xsec2p}), according to shell-model calculations. The states at 4707 and \SI{5230}{keV} are given a $(2,4)^+$ assignment, because these states decay only to the $4^+_1$ \SIhyp{2322}{keV} state. The remaining five states, i.e., those at \num{2288}, \num{2846}, \num{3678}, \num{4153}, and \SI{4919}{keV} are given a $(0,2,4)^+$ assignment. In the next section, the momentum distribution analysis for the two-proton knockout reaction is utilized to further constrain the spin-parity of these states.

\begin{figure}[tb]
\includegraphics[scale=0.5]{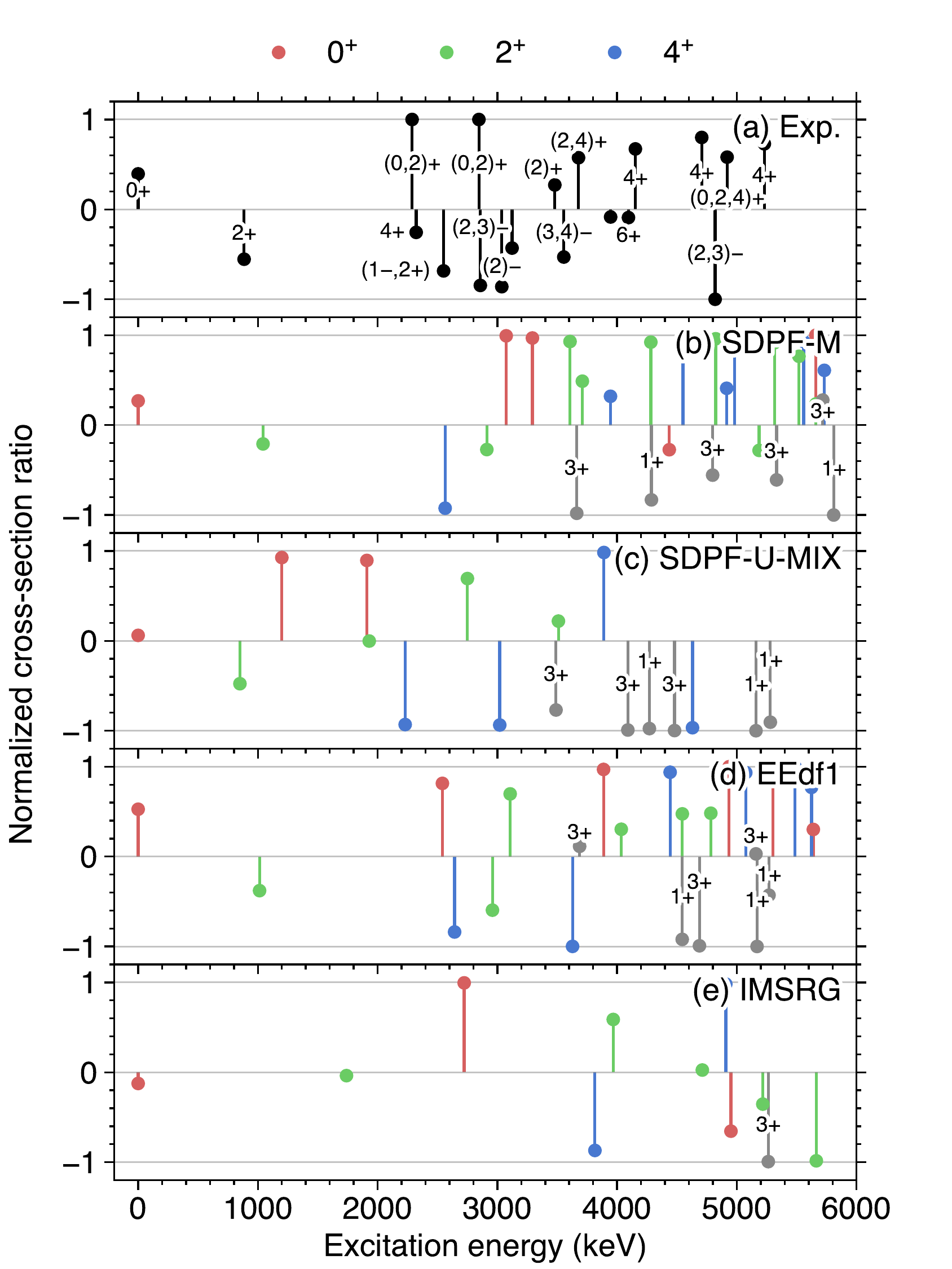}
\caption{Differential normalized cross-section ratios defined in Eq.~(\ref{eq:ratio}). The experimental values (a) are compared with those taken from eikonal-based reaction calculations coupled with shell-model overlaps from (b) SDPF-M, (c) SDPF-U-MIX, (d) EEdf1, and (e) IMSRG. For simplicity, negative-parity states are omitted. Ratios of $+1$ ($-1$) correspond to dominant observation in two-proton (one-neutron) knockout.}
\label{fig:xsecratio}
\end{figure}

To make the above discussion more quantitative, the experimental and theoretical cross sections were analyzed following the approach of Ref.~\cite{ELM19}. In Fig.~\ref{fig:xsecratio}, the vertical axis indicates the values of the differential normalized cross-section ratio defined as
\begin{equation}
\label{eq:ratio}
r = \frac{R_{2p}-R_{1n}}{R_{2p}+R_{1n}}
\end{equation}
where $R_{1n}=\sigma_{1n}(J^\pi_f)/\sigma_{1n}^\mathrm{incl}$ and $R_{2p}=\sigma_{2p}(J^\pi_f)/\sigma_{2p}^\mathrm{incl}$. Therefore, $r=\pm 1$ means exclusive population in either of the reactions. Theoretical calculations predict that high-lying $0^+$, $2^+$, and $4^+$ states are generally strongly populated in two-proton knockout, resulting in large positive values of the ratio. This is a validation of the spin-parity assignments made in the present data analysis. In addition to the seven states discussed above, the \SI{3480}{keV} state shows a positive value, and this state is given a $(0,2,4)^+$ assignment at this stage. Although not explicitly shown in Fig.~\ref{fig:xsecratio}, theoretical cross sections populating negative-parity states in the two-proton knockout reaction are expected to be negligibly small, or such levels are outside the model space.

\subsection{Momentum distributions}

\subsubsection{One-neutron knockout reaction}

\begin{figure}[tb]
\includegraphics[scale=0.5]{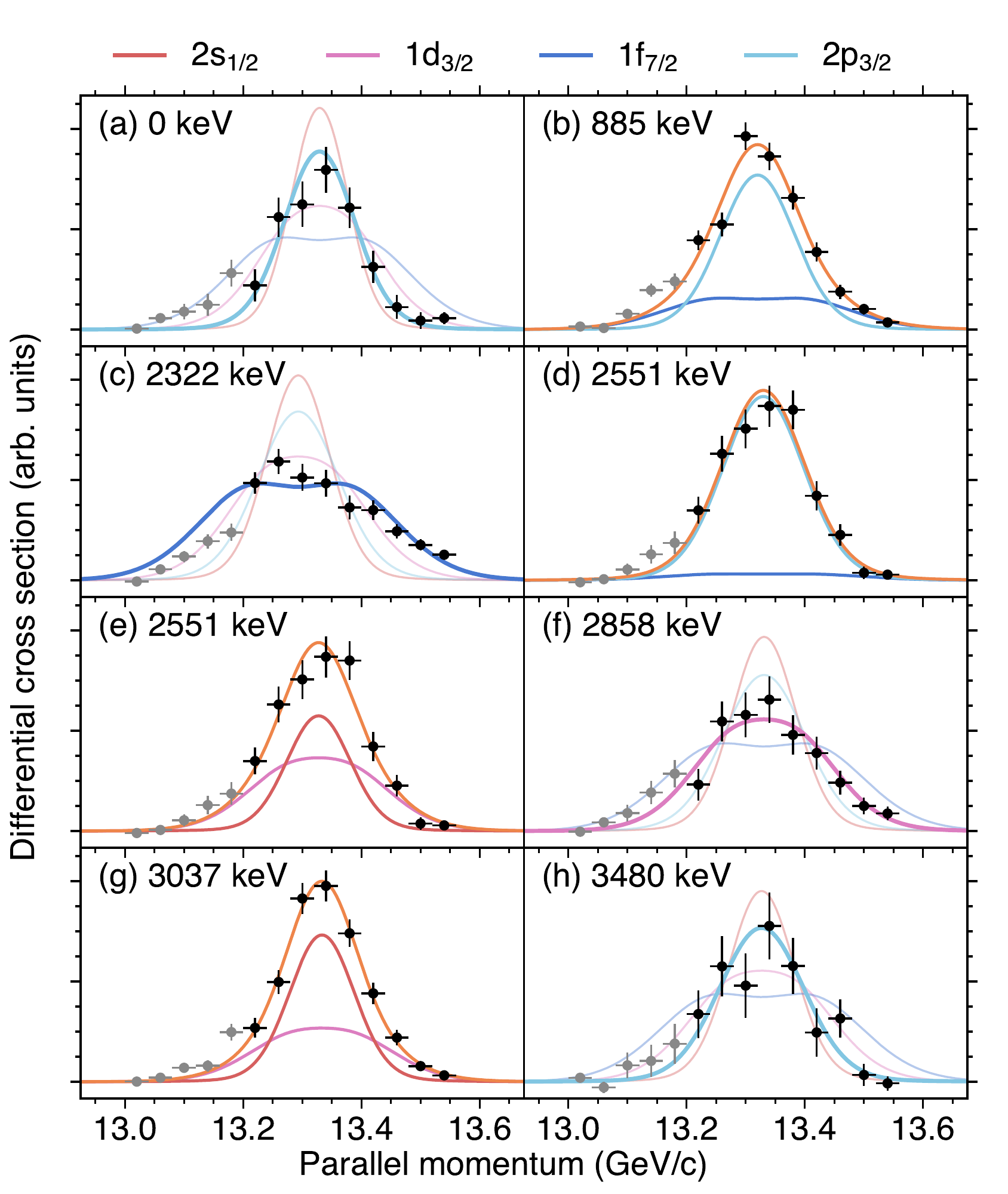}
\caption{Momentum distributions observed in the one-neutron knockout reactions from $^{33}$Mg. For the states at (b) \SI{885}{keV}, (d,e) \SI{2551}{keV}, and (g) \SI{3037}{keV}, the experimental distributions are fitted by a combination of two different orbitals. The orange curve shows the sum of these two contributions. For the \SIhyp{2551}{keV} state, two possibilities of having two different $nlj$ contributions, i.e., $(1f_{7/2},2p_{3/2})$ and $(1d_{3/2},2s_{1/2})$, respectively corresponding to the $2^+$ and $1^-$ spin-parity assignments, are considered.}
\label{fig:momdist1n}
\end{figure}

The momentum distributions associated with individual final states were extracted in a similar manner to the exclusive cross sections. In Fig.~\ref{fig:momdist1n}, experimental one-neutron knockout momentum distributions are compared with calculations. The theoretical distributions are folded with the experimental resolution. As the low-momentum side of the experimental distribution receives contributions from both indirect processes and more dissipative collisions, the normalization of the theoretical curves was determined using the area under the black data points.

As can be seen in Fig.~\ref{fig:momdist1n}(a), the ground-state momentum distribution is compatible with the $2p_{3/2}$ calculation. This observation directly implies that the spin-parity of the $^{33}$Mg ground-state is $3/2^-$. This ground-state spin-parity has been a controversial topic~\cite{NEY11} over the past decades; a magnetic-moment measurement suggested a $3/2^-$ assignment~\cite{YOR07}, while $\beta$-decay studies proposed a spin-parity of $3/2^+$~\cite{NUM01,TRI08b}. Although a $3/2^-$ assignment was proposed again in the one-neutron knockout reaction from $^{34}$Mg~\cite{BAZ21} and the discussion is now converging, the present observation fortifies this claim. An important aspect of the negative-parity assignment is that the $^{33}$Mg ground state is now considered to be of the 3p2h nature. This is evidenced by the large cross sections associated with the $0^+_1$ and $4^+_1$ states, which imply a sizable neutron occupation in the $2p_{3/2}$ and $1f_{7/2}$ orbitals, respectively.

Following the establishment of the $^{33}$Mg ground-state spin-parity, the possible orbitals for the neutron removal for the population of the $2^+$ \SIhyp{885}{keV} state are the $1f_{7/2}$ and $2p_{3/2}$ orbitals. As demonstrated in Fig.~\ref{fig:momdist1n}(b), the extracted momentum distribution can be fitted with a combination of these two orbitals. The $4^+$ \SIhyp{2322}{keV} state should be populated by pure $1f_{7/2}$ neutron knockout. The extracted distribution presents a broad shape and is compatible with $1f_{7/2}$. If the ground-state spin-parity were $3/2^+$, the direct population of a $4^+$ state would require removal of a $1d_{5/2}$ neutron. This scenario is incompatible with the observation, as the removal of such a deeply bound neutron would populate $^{32}$Mg residue at a very high excitation energy with a small cross section.

The \SIhyp{2551}{keV} state is strongly populated in one-neutron knockout, making it a candidate for a negative-parity state. A $1^-$ assignment is suggested, as this state decays both to the ground and $2^+_1$ states, and neutron removal from both $1d_{3/2}$ and $2s_{1/2}$ orbitals contributes to the momentum distribution [see Fig.~\ref{fig:momdist1n}(e)].  However, a $2^+$ assignment to this state is also possible~\cite{TRI08a}. In this case, the momentum distribution should be characterized by a combination of $1f_{7/2}$ and $2p_{3/2}$ components. As can be seen in Fig.~\ref{fig:momdist1n}(d), the distribution is characterized by an almost pure $2p_{3/2}$ contribution. To summarize, the observation is compatible with both $1^-$ and $2^+$ assignments for the \SIhyp{2551}{keV} state. The momentum distribution shown in Figs.~\ref{fig:momdist1n}(d) and (e) does not serve as a discriminator for the spin-parity assignment.

The \SIhyp{3037}{keV} state is also strongly populated in one-neutron knockout and is a negative-parity candidate. This interpretation is in line with the $\beta$-decay result~\cite{TRI08a}. A spin-parity assignment of $2^-$ or $3^-$ is likely, considering the fact that this state exclusively decays to the $2^+_1$ state [see Fig.~\ref{fig:gamgam}(e)] and a direct ground-state decay branch was unobserved. Furthermore, combining the momentum distribution, the possibility of having $3^-$ can be rejected, because the momentum distribution is significantly narrower than what is expected for a pure $1d_{3/2}$ knockout, indicating the presence of a sizable contribution from $2s_{1/2}$; the spin-coupling with the $3/2^-$ $^{33}$Mg ground state and $2s_{1/2}$ does not allow a $3^-$ state to be made, and therefore, a $(2)^-$ assignment is proposed. In Fig.~\ref{fig:momdist1n}(g), the extracted momentum distribution is fitted by a combination of these two orbitals.

Similarly, the state at \SI{2858}{keV} is strongly populated in one-neutron knockout and this state was found to exclusively decay to the $2^+_1$ state as well. In this work, a $(2,3)^-$ assignment is proposed, which is consistent with the earlier $\beta$-decay result~\cite{TRI08a}. The momentum distribution is shown to agree with $1d_{3/2}$ neutron removal, supporting this assignment, but a limit on a $2s_{1/2}$ contribution could not be established because of the statistics.

The \SIhyp{3480}{keV} state is moderately populated both in one-neutron and two-proton knockout. Combining an indication from the two-proton knockout momentum distribution, this state is assigned as $(2)^+$ (discussed in the next section). This state may correspond to the \SIhyp{3488}{keV} state in Ref.~\cite{TAK09} where a tentative assignment of $(1^-,2^+)$ was made. The \SIhyp{3123}{keV} state is also populated in both reactions, but no firm conclusion about $J^\pi$ can be made within the present work. Nevertheless, a \SIhyp{3115}{keV} state with a tentative assignment of $(3^-,4^+)$, reported in Ref.~\cite{TAK09}, could correspond to this state.

The remaining negative-parity candidates, \num{3555}- and \SIhyp{4819}{keV} states~\cite{TRI08a}, are populated almost exclusively in one-neutron knockout. For the \SIhyp{3555}{keV} state, a $(3,4)^-$ assignment was previously given~\cite{TRI08a} and is adopted in the present work. Similarly, a $(2,3)^-$ assignment to the \SIhyp{4819}{keV} state is adopted. Because of the limited statistics, the momentum distributions could not be reliably extracted.

\begin{figure}[tb]
\includegraphics[scale=0.5]{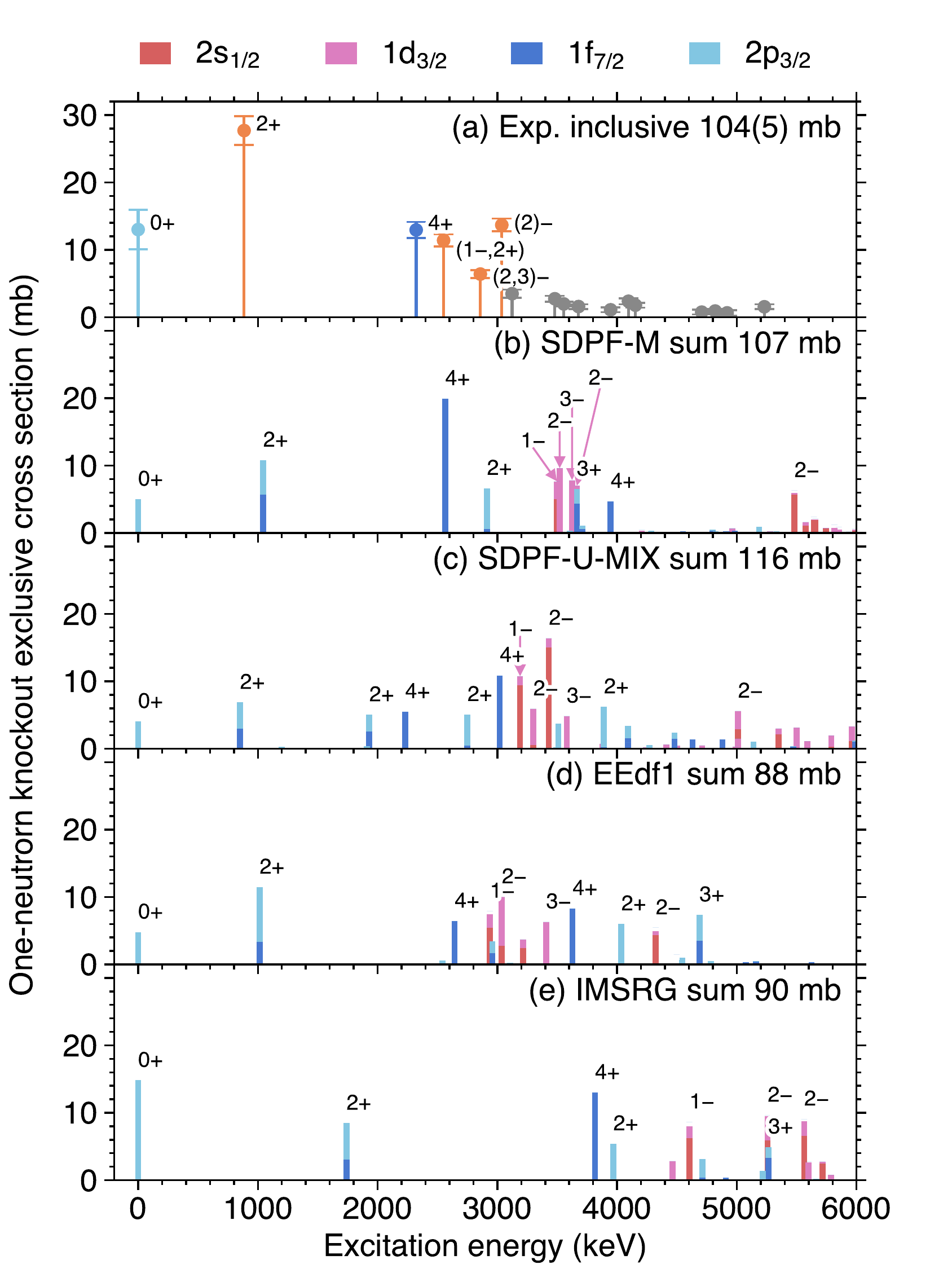}
\caption{Experimental one-neutron knockout cross sections in comparison with calculations. The colors represent contributions of different single-particle orbitals---red: $2s_{1/2}$, cyan: $2p_{3/2}$, pink: $1d_{3/2}$, and blue: $1f_{7/2}$ ($1d_{5/2}$ components are omitted in these plots). The orange bars in panel (a) mean that different single-particle orbitals contribute to their cross sections. The compositions are presented in Fig.~\ref{fig:momdist1n} and Table~\ref{tab:values}.}
\label{fig:xsec1n}
\end{figure}

Based on the spin-parity assignments and the composition of different $nlj$ contributions inferred from the momentum distributions, experimental spectroscopic factors ($C^2S_\mathrm{exp} = \sigma_{1n}/\sigma_\mathrm{sp}$) were calculated. The results are tabulated in Table~\ref{tab:values}.

\subsubsection{Two-proton knockout reaction}

\begin{figure}[tb]
\includegraphics[scale=0.5]{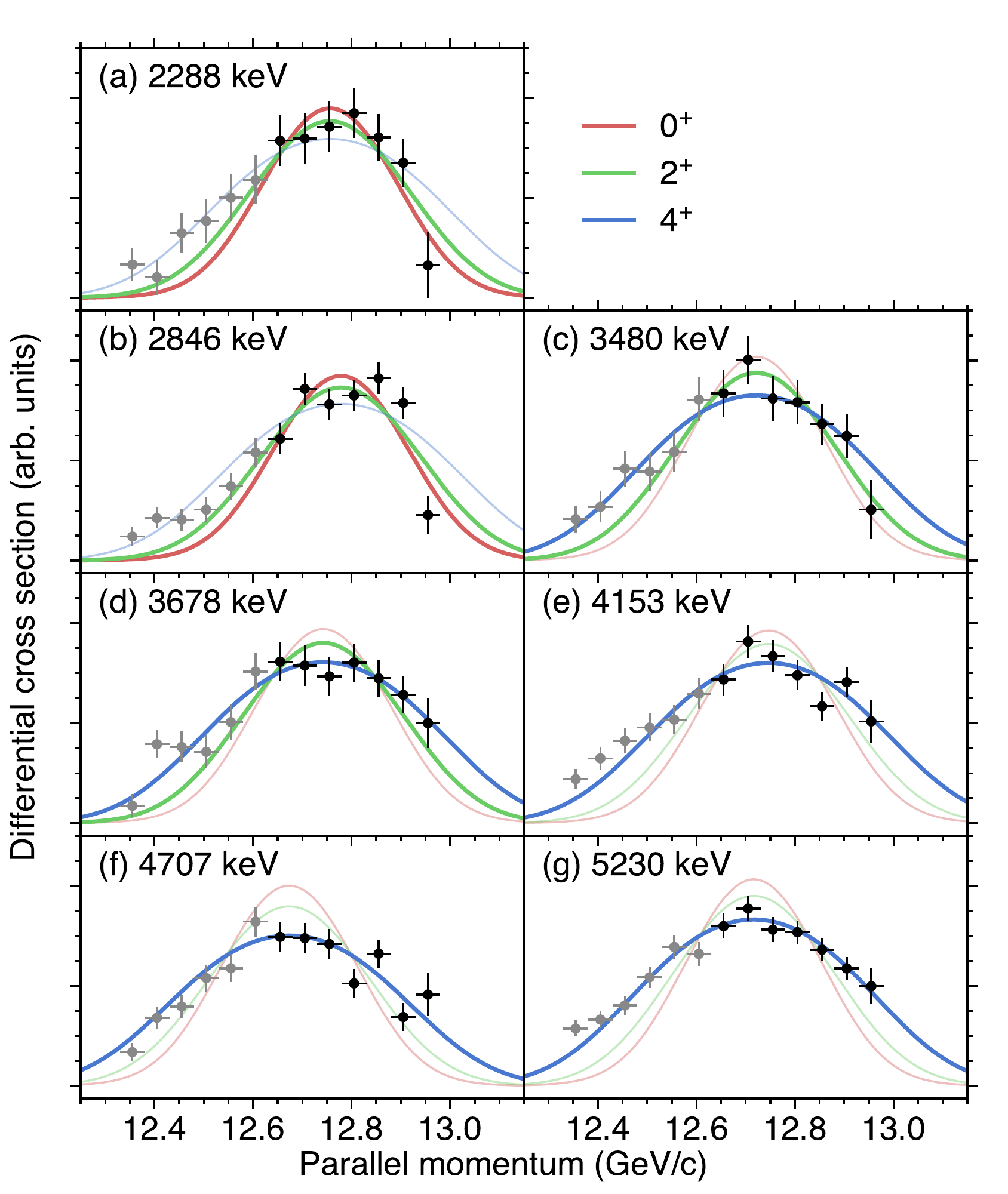}
\caption{Momentum distributions observed in the two-proton knockout reaction from $^{34}$Si. The theoretical curves calculated for final-state spin-parities of $0^+$, $2^+$, and $4^+$, respectively shown by red, green, and blue solid lines, have been normalized by the area under the black data points. TNAs from SDPF-M shell-model calculations were used.}
\label{fig:momdist2p}
\end{figure}

Guided by the differential normalized cross-section ratios, tentative $(0,2,4)^+$ assignments have been made for states with strong population in the two-proton knockout reaction. The momentum distributions in this reaction can be used to further constrain spin-parity assignments for these states. In Fig.~\ref{fig:momdist2p}, extracted momentum distributions are compared with reaction model calculations. In the present measurement, the momentum bite of the S800 spectrograph did not cover the entire momentum distribution of the two-proton knockout reaction. This is partly attributed to the thick reaction target, which is the leading source of the broadening of the distribution. The experimental distributions have been corrected for the limited acceptance. The theoretical momentum distributions, presented in Fig.~\ref{fig:momdist2p}, were calculated using TNAs obtained from SDPF-M shell-model calculations for representative states. Although the momentum distribution depends on the TNAs, variations in the shape of the distribution are much smaller than the differences between different final-state spins.

The momentum distributions associated with the \num{4153}-, \num{4707}-, \SIhyp{5230}{keV} states, shown in Figs.~\ref{fig:momdist2p}(e), (f), and (g), agree well with the calculated $4^+$ momentum distributions, and therefore, these states are firmly assigned as $4^+$. In particular, the \SIhyp{5230}{keV} momentum distribution serves as a clean test case, because this level is lying at a high excitation energy and no $\gamma$ rays feeding this state were observed. The \SIhyp{3480}{keV} state is populated in both one-neutron knockout and two-proton knockout reactions with moderate cross sections. The two-proton knockout momentum distribution is compatible with $2^+$ [see Fig.~\ref{fig:momdist2p}(c)]. A $4^+$ assignment is in contradiction with the narrow momentum distribution observed for the one-neutron knockout reaction shown in Fig.~\ref{fig:momdist1n}(h). A direct population of a $4^+$ state would result in a pure $1f_{7/2}$ momentum distribution, which is much wider than the observation. Therefore, a $(2)^+$ assignment is proposed here. The momentum distribution for the \SIhyp{3678}{keV} state, shown in Fig.~\ref{fig:momdist2p}(d), is not very conclusive, and a $(2,4)^+$ assignment is given in the present work. In the previous section, the \SIhyp{4919}{keV} state was tentatively assigned as $(0,2,4)^+$. Due to the limited statistics, the spin-parity for this state cannot be further constrained.

The \num{2288}- and \SIhyp{2846}{keV} states were, like other cases, tentatively assigned as $(0,2,4)^+$. In Figs.~\ref{fig:momdist2p}(a) and (b), it can be seen that the momentum distributions are significantly sharper than the others, and thus the possibility of $4^+$ assignments can clearly be rejected. However, these distributions show tail structures on the low-momentum side, making the distinction between $0^+$ and $2^+$ unclear. Presently, $(0,2)^+$ assignments are proposed for these two states. Based on the excitation energy and the absence of the ground-state decay, either of these potentially correspond to the $0^+_3$ state predicted at \SI{2.22}{MeV} in a phenomenological three-level mixing approach presented in Ref.~\cite{MAC16}.

A future experiment of one-proton knockout from $^{33}$Al could shed more light on the spin-parity of the states observed in the present two-proton knockout reaction, although the level population could be complicated by the proximity to the island of inversion.

\section{Discussion}
With the new experimental information obtained in the present measurement, the structure of $^{32}$Mg can be discussed in detail. In this section, full comparisons with shell-model calculations, intuitive illustrations of the rich structure coexisting in $^{32}$Mg, and the updated systematics along the $N=20$ isotones are presented.

\subsection{Comparison to shell-model calculations}
\label{sect:comparisontosm}

\subsubsection{Level energies}

To get further insight into the theoretical description of $^{32}$Mg, large-scale shell-model calculations have been performed using the interactions described in Sec.~\ref{sect:structuralcalc}. In Fig.~\ref{fig:levelscheme}, one can see that all of the interactions, except IMSRG, well reproduce the level energies for the $2^+_1$ and $4^+_1$ states in $^{32}$Mg. However, the reproduction of the experimental excitation energy for the $0^+_2$ state is a challenge. As presented in Table~\ref{tab:excalc}, the excitation energies predicted by SDPF-M and EEdf1 are much higher than the experimental value of \SI{1058}{keV}~\cite{WIM10}. Only SDPF-U-MIX gives an accurate prediction for the $0^+_2$ energy.

\begin{table}[tb]
\centering
\caption{Comparison of experimental and theoretical excitation energies for the $2^+_1$ and $0^+_2$ states in $^{32}$Mg and $^{34}$Si. The numbers are given in units of \si{MeV}. The experimental values were taken from Refs.~\cite{WIM10,ROT12}.}
\label{tab:excalc}
\begin{tabular}{lSSSSS}
\toprule
{$^{32}$Mg} & {Exp.} & {SDPF-M} & {SDPF-U-MIX} & {EEdf1} & {IMSRG}  \\
\colrule
$E_\mathrm{x}(2^+_1)$ & 0.885 & 1.04 & 0.85 & 1.01 & 1.77 \\
$E_\mathrm{x}(0^+_2)$ & 1.058 & 3.07 & 1.20 & 2.54 & 2.64 \\
\colrule
{$^{34}$Si} & {Exp.} & {SDPF-M} & {SDPF-U-MIX} & {EEdf1} & {IMSRG}  \\
\colrule
$E_\mathrm{x}(2^+_1)$ & 3.326 & 2.60 & 3.45 & 2.45 & 4.45 \\
$E_\mathrm{x}(0^+_2)$ & 2.719 & 1.82 & 2.58 & 2.44 & 3.55 \\
\botrule
\end{tabular}
\end{table}

In the present work, the lowest state with a negative-parity assignment was established at \SI{2858}{keV}, and this observation is somewhat better reproduced by the EEdf1 calculation. In the IMSRG calculations, excitation energies are generally overestimated, as was already pointed out in Ref.~\cite{MIY20}. However, this also applies to neighboring isotopes and thus the trend along the Mg isotopic chain and the structural evolution can still be discussed in a meaningful way. It is worth noting that the $0^+_2$ state is predicted to be closer in energy to the $2^+_1$ state, unlike the SDPF-M and EEdf1 calculations.

Here, we compare the available observables for $^{34}$Si to further benchmark the calculations. The experimental and theoretical excitation energies for the $2^+_1$ and $0^+_2$ states are summarized in Table~\ref{tab:excalc}. The experimental values are very well reproduced by the SDPF-U-MIX calculations, whereas SDPF-M and EEdf1 show reduced excitation energies, pointing to less doubly-magic features and possibly underestimated effective gap sizes between the $sd$ and $pf$ orbitals.

\subsubsection{One-neutron knockout cross sections}

For the calculation of exclusive cross sections, not only the final states, but also the initial-state wave function affects the spectroscopic factors. All of the interactions, i.e., SDPF-M, EEdf1, SDPF-U-MIX, and IMSRG, predict a $3/2^-$ as the $^{33}$Mg ground state~\cite{BAZ21}, providing a first test. 

In Fig.~\ref{fig:xsec1n}, the theoretical one-neutron knockout cross sections, calculated by Eq.~(\ref{eq:theoxsec}), are displayed together with the experimental cross sections. In the model space of the present work, for all negative-parity states with $J=1\text{--}4$, the $1d_{5/2}$ orbital contributes to the one-neutron knockout cross section. However, such cross sections are predicted to be very small, with spectroscopic factors much less than \SI{10}{\percent} of those for the $1d_{3/2}$ orbital, and these are omitted in Fig.~\ref{fig:xsec1n}. The observed large $4^+$ cross section, and thus the large $1f_{7/2}$ strength, is best reproduced with the SDPF-M interaction. In general, the SDPF-U-MIX and EEdf1 results show more fragmented strengths for positive-parity states. For the $2^+_1$ state, the experimental cross section is larger than the predictions, at least, by a factor of 2. Even though the $2^+_1$ state is expected to serve as a collector of many weak transitions and the cross section may potentially be overestimated, the contribution of $1f_{7/2}$ and $2p_{3/2}$ orbitals, inferred from the momentum distribution, agrees with the calculations (see Fig.~\ref{fig:momdist1n} and Table~\ref{tab:values}). The cross sections calculated with the IMSRG spectroscopic factors are similar to those calculated using the shell-model results and provide a reasonable reproduction of the experimental values.

According to the Nilsson-model based analysis presented in Ref.~\cite{MAC17}, the spectroscopic factors for the $0^+_1$, $2^+_1$, and $4^+_1$ states are predicted to be $C^2S(2p_{3/2}) = 0.24$, $[C^2S(2p_{3/2}),C^2S(1f_{7/2})] = [0.24,0.18]$, and $C^2S(1f_{7/2}) = 0.33$, respectively, calculated with empirically adjusted Nilsson wave functions. The values for $2^+_1$ and $4^+_1$ are about half of the experimental spectroscopic factors, and the reason for this difference remains to be understood.

\subsubsection{Two-proton knockout cross sections}

\begin{figure}[tb]
\includegraphics[scale=0.5]{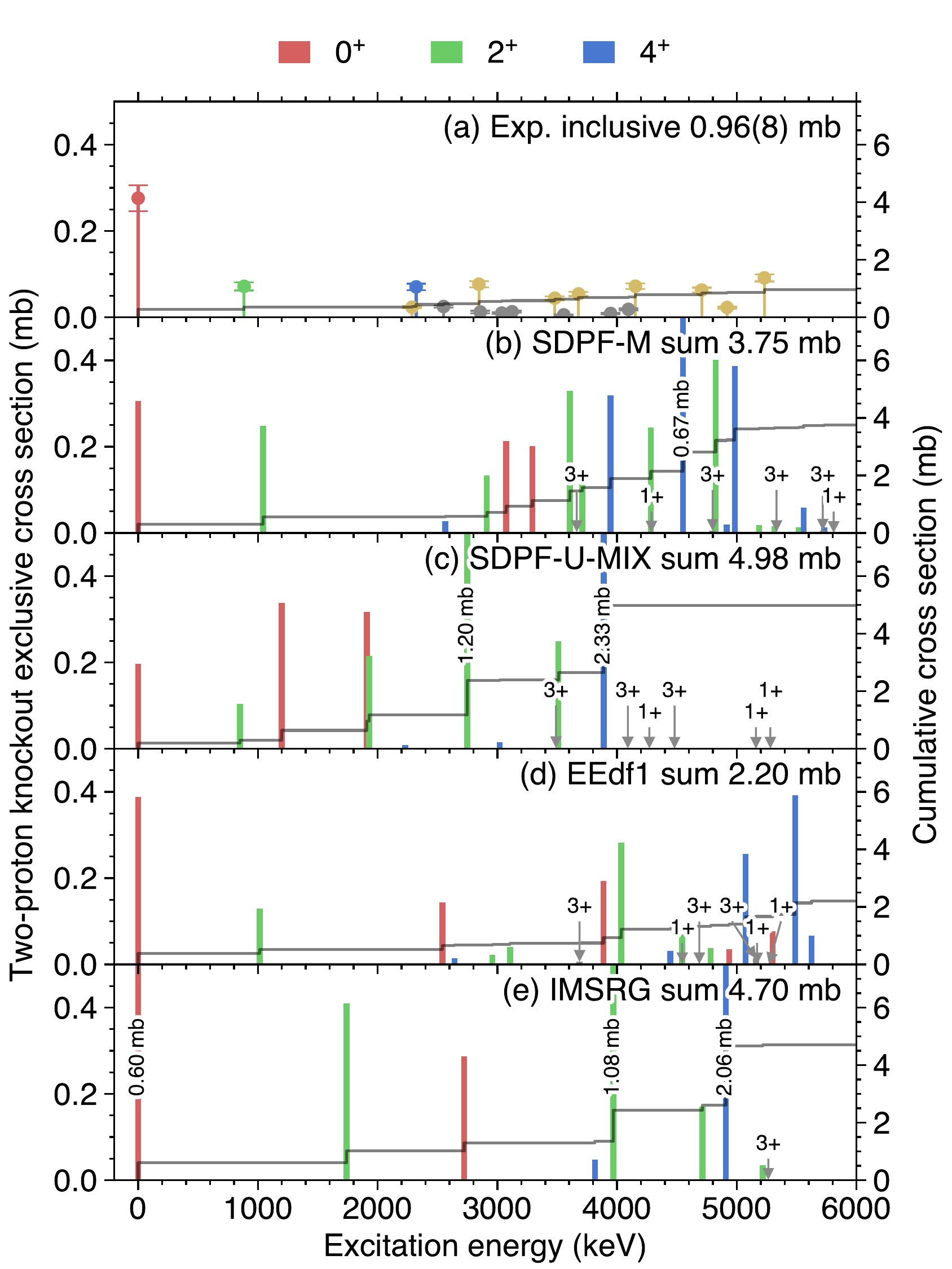}
\caption{Two-proton knockout cross sections in comparison with calculations. The bars are colored according to their spin-parity---$0^+$: red, $2^+$: green, and $4^+$: blue. The yellow bars in panel (a) represent candidates for either $0^+$, $2^+$, or $4^+$. The running sums of the exclusive cross sections are indicated as black solid lines (see the right vertical axis for the scale).}
\label{fig:xsec2p}
\end{figure}

In Fig.~\ref{fig:xsec2p}, the measured two-proton knockout cross sections are compared with theoretical predictions. These combine structure information in the form of TNAs with reaction model calculations of the proton-pair removal cross section. For SDPF-M, all positive-parity states with $J=0\text{--}4$ lying below the neutron threshold ($S_n = \SI{5778}{keV}$~\cite{OUE11}) have been calculated. Therefore, the sum of these cross sections is interpreted as the theoretical inclusive cross section. For EEdf1 (SDPF-U-MIX), only the six (three or four) lowest states for each spin have been calculated. Hence, the sum of these cross sections provides a lower limit of the theoretical inclusive cross section.

Remarkably, SDPF-U-MIX predicts very concentrated strengths to bound states, especially to the $2^+_3$ and $4^+_3$ states, and as a result, the theoretical inclusive cross section greatly exceeds the measured one. The SDPF-M result presents more fragmented strengths than SDPF-U-MIX, and population of yrare states is even more hindered in the EEdf1 calculations. The IMSRG result is similar to SDPF-U-MIX in that it shows very concentrated strengths to bound states, but the $2^+_2$ and $4^+_2$ carry the largest cross sections. However, the cross sections are potentially overestimated, because of the use of bare annihilation operators for calculating TNAs. Ideally, as was mentioned in Sec.~\ref{sect:structuralcalc}, these should be replaced with effective operators derived from IMSRG. A reduction factor of around \num{0.5} is expected from the systematics established in the $sd$-shell test cases~\cite{TOS06}. In terms of the inclusive cross section, the EEdf1 result is compatible with the quenching factor of \num{0.5}, but a more significant reduction is needed for the other results to reconcile with the observation. Larger reductions were recently observed in systems further from stability where large structural changes take place~\cite{MUR19}, but a complete understanding is not yet obtained. These findings pose further questions about the theoretical description of the transition into the island of inversion and highlight the importance of detailed spectroscopic information beyond the locations of excited states.

In general, odd-spin cross sections are very small as compared to even-spin states, as was the case for $^{22}$Mg in Ref.~\cite{LON20}. For SDPF-M and EEdf1, the proton $sdpf$ model space allows the population of negative-parity and $J=5^+$ states in the two-proton knockout reaction.  However, since the occupation of the proton $pf$ orbitals in the ground state of $^{34}$Si is very small, the resulting cross sections to such states are negligible. It is possible that negative-parity states are formed by leaving a proton hole in $1p$ orbitals, but such configurations are outside the present model space and such strength is expected only at very high excitation energies.

The underlying $jj$-basis TNAs, used for the cross-section calculations, are also visualized in Fig.~\ref{fig:hwcomposition} for the lowest three $0^+$, $2^+$, and $4^+$ states. It can be seen that the resulting cross section is correlated with the magnitude of the $(1d_{5/2})^2$ amplitudes. In the SDPF-U-MIX calculations, the very large cross sections populating the $2^+_3$ and $4^+_3$ are attributed to the substantial wave-function overlaps creating two proton holes in the $1d_{5/2}$ orbital. The same reasoning applies to the $2^+_2$ and $4^+_2$ in IMSRG.

\subsubsection{Configurations in the shell-model wave functions}

\begin{figure}[tb]
\includegraphics[scale=0.5]{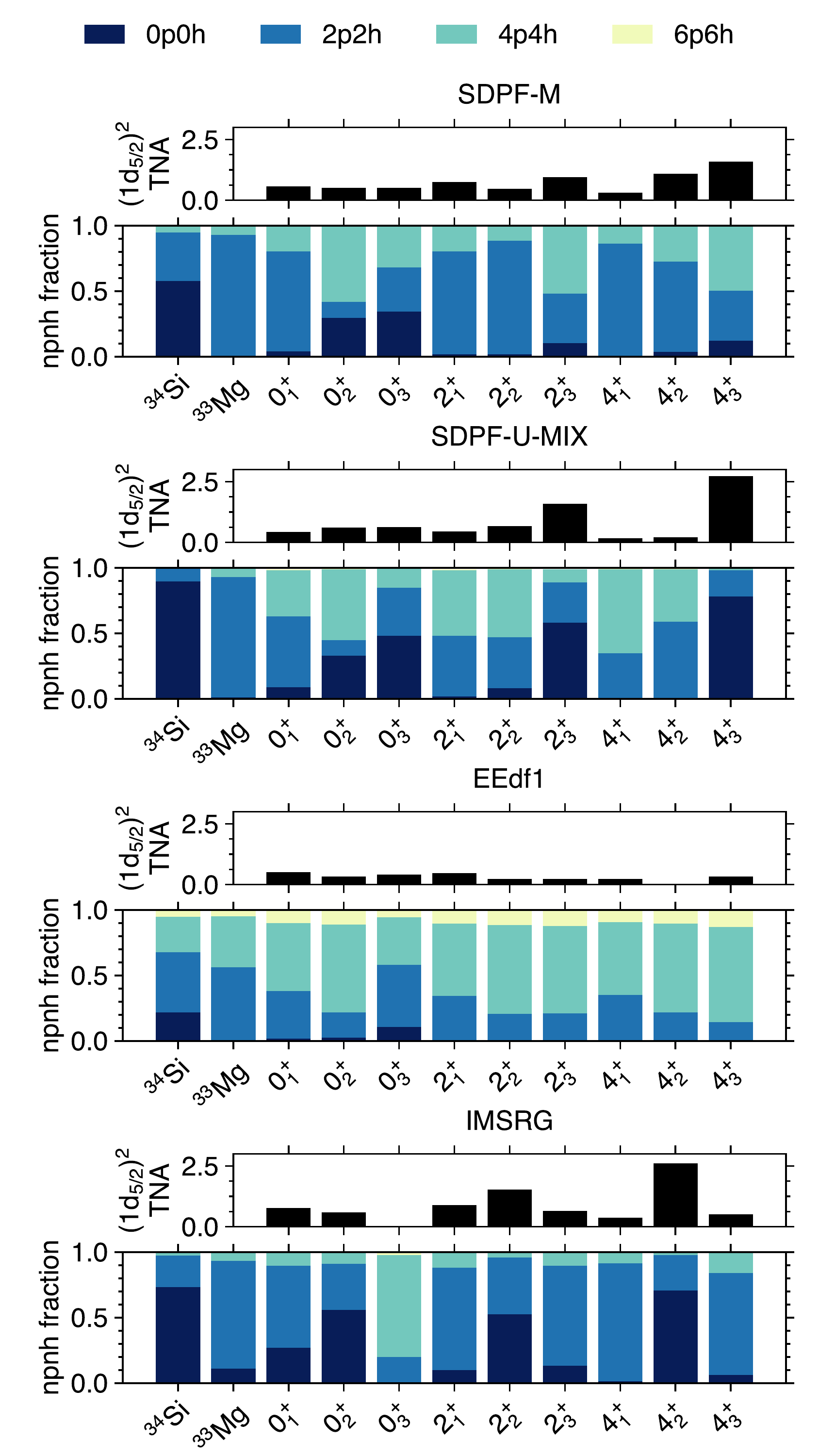}
\caption{Composition of $n$p$n$h excitations in shell-model wave functions. The $(1d_{5/2})^2$ component of TNAs, calculated in the $jj$ basis, are shown together for the lowest three $0^+$, $2^+$, and $4^+$ states in $^{32}$Mg.}
\label{fig:hwcomposition}
\end{figure}

It is worthwhile to investigate the compositions of shell-model wave functions to understand the underlying configuration in each state. To this end, fractions of $n$p$n$h excitations in the lowest three $0^+$, $2^+$, and $4^+$ states of $^{32}$Mg are visualized in Fig.~\ref{fig:hwcomposition}.

The different interactions result in distinct wave function compositions. Interestingly, in SDPF-M and EEdf1 calculations, the $^{32}$Mg states presented in Fig.~\ref{fig:hwcomposition} are found to be of predominantly intruder nature. While the $0^+_2$ and $0^+_3$ states are dominated by particle-hole excitations in the SDPF-M and EEdf1 calculations, the $0^+_3$, $2^+_3$ and $4^+_3$ states in the SDPF-U-MIX calculations have large (typically greater than \SI{50}{\percent}) 0p0h contributions, thus closer to the na\"ive expectation that the normal, spherical configuration is coexisting with deformed configurations in $^{32}$Mg. In SDPF-U-MIX, the $0^+_2$ is described as a hybrid of spherical 0p0h and super-deformed 4p4h configurations, as pointed out in Ref.~\cite{CAU14}. This composition is found to be very similar to that in SDPF-M. The IMSRG calculations are different in that the $0^+_2$, $2^+_2$ and $4^+_2$ states are characterized by large 0p0h fractions. It is worth noting that, in the phenomenological three-level mixing model in Ref.~\cite{MAC16}, the $0^+_3$ state has a large fraction of the 0p0h component, presenting a similar picture as SDPF-U-MIX.

For $^{32}$Mg, all of the calculations show a sizable amount of 4p4h mixing, which was recently proposed to play an essential role in describing the shape mixing, as well as the transition into the island of inversion~\cite{CAU14,MAC16}. The EEdf1 wave functions generally have more 4p4h contributions to all the states than the other calculations. As discussed in Ref.~\cite{TSU17}, this interaction tends to promote $n$p$n$h excitations. Nevertheless, this does not necessarily mean strong quadrupole deformation, since not only quadrupole correlation but also other types of correlations, e.g., pairing correlation, can promote excitations of this kind. Although not shown in Fig.~\ref{fig:hwcomposition}, in the SDPF-M, SDPF-U-MIX, and EEdf1 calculations, the lowest $1^-$, $2^-$, and $3^-$ states are dominated by intruder, 3p3h and 5p5h configurations. Therefore, normal 1p1h configurations such as those observed in $^{30}$Mg~\cite{KIT20} are not predicted as the lowest-lying negative-parity states in the island-of-inversion nucleus $^{32}$Mg. The negative-parity states in the IMSRG calculations tend to show more mixed 1p1h and 3p3h configurations. Nevertheless, several states dominated by 3p3h configurations are found below the neutron threshold.

Compositions of $n$p$n$h excitations in the ground states of $^{33}$Mg and $^{34}$Si are also displayed in Fig.~\ref{fig:hwcomposition}. In all of the calculations, the $^{33}$Mg ground state is dominated by 3p2h excitations, and this is in line with the experimental findings. The $^{34}$Si ground state is considered to be doubly-magic, as evidenced by various observations. As presented in Table~\ref{tab:excalc}, its high $E_\mathrm{x}(2^+_1)$~\cite{BAU89} and $E_\mathrm{x}(0^+_2)$~\cite{ROT12}, combining the small $B(E2;0^+_1\to2^+_1)$~\cite{IBB98}, are direct indications of shell closure. Additionally, experimental data on neutron and proton occupation in $^{34}$Si~\cite{END02,MUT17,JON20} and neutron single-particle structure in $^{35}$Si~\cite{BUR14} validated this interpretation. In some of the shell-model calculations, however, the 0p0h component is reduced to \SI{60}{\percent} (SDPF-M) and even \SI{20}{\percent} (EEdf1), showing less doubly-magic features. These outcomes are, as stated earlier, already speculated from the theoretical values for $E_\mathrm{x}(2^+_1)$ and $E_\mathrm{x}(0^+_2)$. These results suggest a similarity to the nucleus $^{56}$Ni ($N=28$ and $Z=28$), which is considered as being doubly-magic based on experimental observations, whereas shell-model calculations show a quenched doubly-magic composition in its wave function~\cite{OTS98}. The diminished 0p0h compositions in SDPF-M and EEdf1 are in contrast with a 0p0h wave-function fraction of \SI{90}{\percent} in SDPF-U-MIX. We also point out that the doubly-magic $\pi(1d_{5/2})^6 \nu(1d_{5/2})^6(2s_{1/2})^2(1d_{3/2})^4$ configuration in the $^{34}$Si ground state amounts to \SI{80}{\percent} with this interaction.

\subsection{Coexisting normal and intruder configurations}

As discussed earlier in Ref.~\cite{KIT21}, the present experimental data allowed us to decipher the rich level structure in $^{32}$Mg. The states can be sorted into three groups according to their underlying configurations, as displayed in different colors in Fig.~\ref{fig:levelscheme}.

\begin{enumerate}
\item Deformed intruder ground-state band. Following the first observation of the $6^+_1$ state in Ref.~\cite{CRA16}, the yrast ground-state band in $^{32}$Mg was interpreted as strongly deformed, guided by the level spacing. Mixing of a large fraction of $n$p$n$h configurations is essential to drive the ground state deformation. Markedly, the excitation energies for the $2^+_1$, $4^+_1$, and $6^+_1$ states are accurately reproduced by the SDPF-U-MIX calculation, where 2p2h and 4p4h configurations dominate over 0p0h by \SI{90}{\percent}.
\item Intruder 3p3h negative-parity states. The negative-parity states populated in the one-neutron knockout reaction should be characterized by 3p3h configurations, whose structure is different from non-intruder 1p1h negative-parity states observed, e.g., in $^{30}$Mg~\cite{KIT20}. The 3p3h interpretation is supported by the $^{33}$Mg ground-state structure, which has shown to be dominated by an intruder-dominated 3p2h configuration, and also the shell-model calculations for negative-parity states in $^{32}$Mg.
\item Normal 0p0h positive-parity states. In the two-proton knockout reaction, $0^+$, $2^+$, and $4^+$ states were strongly populated. As the direct two-proton removal, to first order, leaves the neutron configurations untouched, these states are formed by excitations only in the proton side. Combining the doubly-magic nature of $^{34}$Si, these states are considered to be the members of the off-yrast normal-configuration bands, which have not been explored experimentally. In the present measurement, candidates for the $0^+_3$ state were identified at 2288 and \SI{2846}{keV}. The predicted $0^+_3$ state in Ref.~\cite{MAC16} exhibits the largest 0p0h contribution among the three $0^+$ states. Additionally, from a simplified perspective, the $0_2^+$ state is understood as a shape-coexisting, normal state, serving as the counterpart of the $^{34}$Si ground state. This is in line with the fact that the $0_2^+$ state was observed only in the $t$($^{30}$Mg,$p$)~\cite{WIM10} and the two-proton knockout reaction from $^{34}$Si~\cite{ELD19}. In both cases, initial states are considered to be non-intruder, and thus the population of normal states is favored.
\end{enumerate}

\subsection{Systematics}

\begin{figure}[tb]
\includegraphics[scale=0.5]{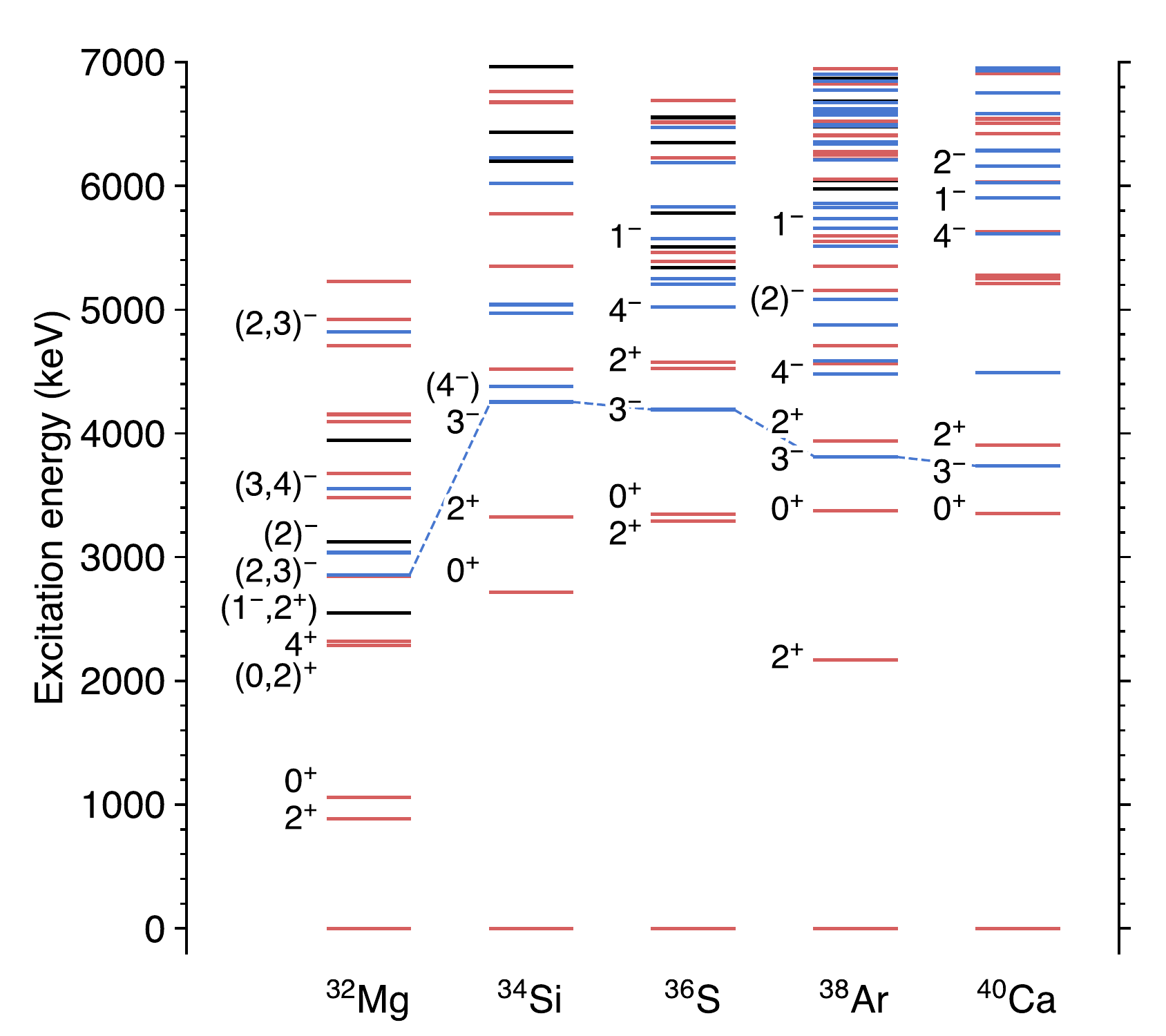}
\caption{Level systematics of the $N=20$ isotones. The levels in $^{32}$Mg are taken from the present work, while those in $^{34}$Si are from Ref.~\cite{LIC19}. The others are adopted from the latest ENSDF database as of this writing. Positive-parity (negative-parity) states are indicated by red (blue) horizontal lines. The lowest $3^-$ states are connected by dashed lines.}
\label{fig:systematics}
\end{figure}

The systematic behavior of excited levels in the $N=20$ isotones is shown in Fig.~\ref{fig:systematics}. While the locations of negative-parity states down to $^{34}$Si were discussed previously in Ref.~\cite{LIC19}, the present work established the lowest firmly-assigned negative-parity state in $^{32}$Mg at \SI{2858}{keV}, further extending the systematics. It can be seen that the lowest negative-parity states remain around \SI{4}{MeV} from $^{40}$Ca to $^{34}$Si, but at $^{32}$Mg it suddenly drops by \SI{1.4}{MeV}. This compares well with the rapid decrease of \SI{2.0}{MeV} observed for the $3^-_1$ states in the $N=18$ chain~\cite{KIT20}, albeit to a slightly lesser degree. To first order, the excitation-energy drop is correlated with the reduction of the effective size of the $N=20$ gap, pointing to the shell evolution approaching the island of inversion. However, the drop is also affected by correlation effects~\cite{CAU14}, unlike the systematics established for the $N=18$ isotones~\cite{KIT20}, where the underlying configuration is understood in terms of simple 1p1h excitations.

Even though a $1^-$ state has not been observed in $^{34}$Si, an extrapolation based on the trend from $^{40}$Ca to $^{36}$S would place it at around \SI{5}{MeV}. Assuming $J^\pi=1^-$ for the \SIhyp{2551}{keV} state, there could be a rapid reduction of $1^-$ states from $^{34}$Si to $^{32}$Mg. All of the calculations predict population of the $1^-_1$ state in $^{32}$Mg with cross sections of around \SI{10}{mb} (see Figs.~\ref{fig:xsec1n}(b--e)). The observed cross section populating the \SI{2551}{keV} state is close to this expectation, but the same argument applies to the $2^+_2$ state, with somewhat smaller theoretical cross sections of around \SI{5}{mb}. In this regard, a more detailed in-beam $\gamma$-ray measurement which utilizes angular distribution and linear polarization information of $\gamma$ rays, as well as a multi-step Coulomb excitation measurement are encouraged. However, these will require a next-generation experimental facility.

\section{Summary and conclusion}

States in $^{32}$Mg have been studied by means of high-statistics in-beam $\gamma$-ray spectroscopy through two different direct-reaction probes, one-neutron knockout from $^{33}$Mg and two-proton knockout from $^{34}$Si. The first results were presented in Ref.~\cite{KIT21}, and in this work, we covered details of the experimental data and more comprehensive comparisons with theoretical calculations.

The results from one-neutron knockout allowed us to strengthen the $3/2^-$ ground-state spin-parity assignment for $^{33}$Mg, pointing to a 3p2h nature, which has been the subject of a longstanding debate. Because of the intruder nature of $^{33}$Mg, 3p3h negative-parity states in $^{32}$Mg are expected to be selectively populated in one-neutron knockout. Indeed, such states were identified experimentally, and the firmly-assigned lowest-energy negative-parity state was established at \SI{2858}{keV}.  This observation extended the systematics along the $N=20$ isotonic chain, highlighting the shell evolution approaching the island of inversion.

Several positive-parity states with spin-parities of $0^+$, $2^+$, and $4^+$ were exclusively populated in two-proton knockout and observed for the first time. Due to the doubly-magic nature of $^{34}$Si, these levels are interpreted as having normal nature. Also, this work provided the first indication for a candidate for the $0^+_3$ state, which has been theoretically predicted. We emphasize that the two-proton knockout reaction has been shown to give a remarkable selectivity to the states embedded in the off-yrast region.

This work unraveled yet another type of structures coexisting in $^{32}$Mg. The states are classified as, according to their nature, (i) ground-state rotational band, (ii) 3p3h intruder negative-parity levels, and (iii) 0p0h normal positive-parity levels. This finding further extends the shape coexisting picture established so far. The obvious next steps include the observation of the band structure built on top of the normal levels. Also, determination of the interband transition strengths would be an experimental challenge.

The experimental level energies and exclusive cross sections in both reactions were confronted with theoretical calculations that combine the well-established reaction model and shell-model overlaps from four different interactions, SDPF-M, SDPF-U-MIX, EEdf1, and IMSRG. Different aspects of the structure of $^{32}$Mg are captured in some of these models. However, a consistent description is not yet obtained, as evidenced by the discrepancy found between the theoretical and experimental two-proton knockout cross sections, which cannot be explained by the systematic quenching of experimental cross sections observed in $sd$-shell, test-case systems. Of relevance is that the doubly-magic feature of the $^{34}$Si ground state is supported by the SDPF-U-MIX results, while the other interactions show less magicity. Also, it was found that the different interactions illustrate distinct $n$p$n$h compositions in excited states of $^{32}$Mg. To summarize, even with the state-of-the-art shell-model calculations, none is able to fully reproduce the experimental observation, and this poses a further, interesting challenge for theoretical models.

\begin{acknowledgments}
We express our gratitude to the accelerator staff at NSCL for their efforts in beam delivery during the experiment. N.K.\ acknowledges support of the Grant-in-Aid for JSPS Fellows (18J12542) from the Ministry of Education, Culture, Sports, Science, and Technology (MEXT), Japan. K.W.\ acknowledges support from the Ministerio de Ciencia e Innovaci\'on (Spain) through the ``Ram\'on y Cajal'' program RYC-2017-22007. A.P.\ is supported in part by the Ministerio de Ciencia, Innovaci\'on y Universidades (Spain), Severo Ochoa program SEV-2016-0597 and grant PGC-2018-94583. The SDPF-M calculations were enabled by the CNS-RIKEN joint project for large-scale nuclear structure calculations and were performed mainly on the Oakforest-PACS supercomputer. N.S.\ acknowledges support from ``Priority Issue on post-K computer'' (hp190160) and ``Program for Promoting Researches on the Supercomputer Fugaku'' (JPMXP1020200105, hp200130, and hp210165) by JICFuS and MEXT, Japan. The IMSRG calculations were performed with an allocation of computing resources on Cedar at WestGrid and Compute Canada and on the Oak Cluster at TRIUMF managed by the University of British Columbia, Department of Advanced Research Computing. J.A.T.\ acknowledges support from the U.K.\ Science and Technology Facilities Council Grant No.\ ST/L005743/1. This work was supported by the U.S.\ Department of Energy (DOE), Office of Science, Office of Nuclear Physics, under Grant No.\ DE-SC0020451 and by the U.S.\ National Science Foundation (NSF) under Grant No.\ PHY-1306297. GRETINA was funded by the U.S.\ DOE, Office of Science. Operation of the array at NSCL is supported by the U.S.\ NSF under Cooperative Agreement No.\ PHY-1102511 (NSCL) and DOE under Grant No.\ DE-AC02-05CH11231 (LBNL).
\end{acknowledgments}

\end{document}